\documentclass[runningheads]{llncs}

\usepackage{microtype}
\usepackage{hyperref}
\usepackage{todonotes}
\usepackage{pgfplots}
\usetikzlibrary{
	pgfplots.dateplot,
}

\usepackage{graphicx}
\usepackage[printwatermark]{xwatermark}
\usepackage{xcolor}
\usepackage{graphicx}

\newwatermark[allpages,color=red!30,angle=0,scale=0.4,xpos=0,ypos=130]{\textit{Draft -- Not Peer Reviewed -- Open For Comments}}
\newwatermark[allpages,color=red!30,angle=0,scale=0.4,xpos=0,ypos=-130]{\textit{Draft -- Not Peer Reviewed -- Open For Comments}}

\begin{document}

\pagestyle{empty}
\section*{Preface}
In the last years a lot of research has been done, both in academia and industry, regarding blockchain technology. Unfortunately, we found very little research covering what properties a system is required to have to be called a blockchain system and, if they do not fulfill these criteria, what terminology can be used to describe them instead.

A phrase attributed to Socrates is that "the beginning of wisdom is the definition of terms". We believe that, especially in such a fast developing area like blockchain technology, a clear definition of terms is important to gain a commonly shared way to communicate concepts and thoughts. \newline

The following paper is a draft we developed in 2018 during internal discussions concerning this topic. We are discussing what properties we believe a blockchain system is required to have to do justice to this term. Furthermore, we argue that an alternative term is needed to discuss technologies not fulfilling all requirements to be called blockchain but being related to this technology. We believe that the term \textit{Decentralized Consensus Technology} might be both generic and specific enough to cover all the advancement in the area of blockchain technology without excluding possible solutions for challenges in that area. \newline

Between the end of 2018 and the beginning of 2019 we submitted the paper both to a conference covering topics of security, privacy and distributed systems and later (after addressing reviewer feedback) to a journal covering the topic of software engineering. Unfortunately, both times the paper was rejected. The reasons for the rejections can be summarized as follows: 
\begin{itemize}
	\item We didn't meet the expected format and the degree of novelty was deemed too low
	\item We lack a scientific process for collecting the presented concepts and data. E.g. a systematic mapping study would have been preferred.
	\item It is unknown in which direction the technology will develop and it cannot be said whether our proposed terminology covers the complete connotation of blockchain systems.
\end{itemize}

Due to various discussions both inside and outside our research group we decided that the following draft might serve as a good starting point for further discussions in the context of blockchain terminology. We therefore decided to publish it on a publicly accessible platform, regardless of whether it was rejected by the conference and journal. We hope that it helps to develop a commonly shared understanding regarding the properties of such systems and the terminology to describe them. Feel free to use it as a basis for further research and to get in touch with us to discuss this topic.

\newpage

\pagestyle{headings}
\title{Properties of Decentralized Consensus Technology -- Why not every Blockchain is a Blockchain}
\titlerunning{Why not every Blockchain is a Blockchain}
\author{Christopher Ehmke
\and
Florian Blum
\and
Volker Gruhn
}
\authorrunning{C. Ehmke et al.}
\institute{University of Duisburg-Essen, Schuetzenbahn 70, 45127 Essen, Germany
\email{firstname.lastname@uni-due.de}}
\maketitle              
\begin{abstract}
Research in the field of blockchain technology and applications is increasing at a fast pace. %
Although the Bitcoin whitepaper by Nakamoto is already ten years old, the field can still be seen as immature and at an early stage. %
Current research in this area is lacking a commonly shared knowledge and consensus about terms used to describe the technology and its properties. %
At the same time this research is challenging fundamental aspects of the Bitcoin core concept. %
It has to be questioned whether all of these new approaches still adequately could be described as blockchain technology. %
We propose to use the term \textit{Decentralized Consensus Technology} as a general category instead. \textit{Decentralized Consensus Technology} consists of decentralized ledger and non-ledger technologies. Blockchain technology in turn is only one of multiple implementations of the \textit{Decentralized Ledger Technology}. %
Furthermore, we identified three main characteristics of \textit{Decentralized Consensus Technology}: \textit{decentralization}, \textit{trustlessness} and \textit{ability to eventually reach consensus}. Depending on the use case of the specific implementation the following additional properties have to be considered: \textit{privacy}, \textit{participation incentive}, \textit{irreversibility} and \textit{immutability}, \textit{operation purpose}, \textit{confirmation time}, \textit{transaction costs}, \textit{ability to externalize transactions and computations} and \textit{scalability possibilities}.

\end{abstract}

\section{Introduction}
Blockchain technology is an innovation said to be one of the most disruptive inventions of the past decades \cite{panetta:2017}. Although, initially only intended to be a decentralized alternative to the traditional centralized financial currency system \cite{nakamoto_bitcoin:_2008}, it is now used in various different scenarios. Since the Bitcoin white paper presenting the first concepts, later named blockchain technology \cite{nakamoto_bitcoin:_2008}, a lot of research has been done to further improve the concept and implementations. In spite of all these improvements, or maybe even because of them, it is hard to describe how a blockchain system is characterized \cite{bonneau_sok:_2015}. There is little commonly shared knowledge and agreed-upon terminology for the specific properties a system has to provide to be seen as a blockchain technology and which properties of a certain system might prevent such a classification. %
This area of research is growing and in the last year a few publications started to address this issue (cf. \cite{conte_de_leon_blockchain:_2017,xu_taxonomy_2017,wust_you_2017}). %
The problem of having no commonly known and accepted description of the system under examination becomes visible when analyzing the state of research for blockchain applications: Nearly all start with their own introduction and definition of blockchain technology. Often some properties are mentioned used to classify a blockchain project as such. Several papers state for example that \textit{decentralization}, \textit{immutability} of included data and being able to operate without relying on \textit{trust} between the participants are fundamental properties of blockchain technology \cite{nakamoto_bitcoin:_2008,bonneau_sok:_2015}. Furthermore, there is a lot of research regarding the modification of certain properties (e.g. see \cite{ateniese_redactable_2017,sankar_survey_2017,patel_blockchain_2017}) while still claiming the resulting systems to be blockchains. 

Remarkably, even though Nakamotos \cite{nakamoto_bitcoin:_2008} white paper is considered to be origin of the blockchain technology, the term "Blockchain" is not used in their paper. We will show that it is hard to point out which properties characterize blockchain technology, especially when several approaches modify, remove or extend ideas from the initial concept by Nakamoto. During the last years the term \textit{Distributed Ledger Technology} (DLT), respectively \textit{Distributed Ledger System} (DLS), came up to handle these terminology shortcomings \cite{conte_de_leon_blockchain:_2017}. Unfortunately, even these terms do not cover all forms of system that came up in the context of blockchain technology. For example, it has to be discussed whether a distributed computation platform like Ethereum \cite{buterin_ethereum_2014} can be accurately described as a (decentralized) ledger. We propose to use the term \textit{"Decentralized Consensus Technology"} (DCT) (cf. \cite{glaser_beyond_2015}) aiming at describing a wider range of technologies in the context of blockchains without being too specific to cover modified versions of the concept. We will show why the distinction between blockchain technology and the term \textit{Decentralized Consensus Technology} (DCT) is important and should be considered in further research. 

This paper is structured as follows: The next section will give a short introduction to the initial concept presented by Nakamoto, derive specific properties of their approach as a baseline for further comparison and highlight some problems of this initial concept. Section~\ref{lab:sec:research} will give a brief overview of current research topics modifying properties of the initial system. Section~\ref{lab:sec:DCS} covers insights about the definition of the term \textit{DCT} and will show that the question which properties characterize \textit{Decentralized Consensus Technology} depends on the application to be developed. Section~\ref{lab:sec:related:work} will present related work in this context and Section~\ref{lab:sec:conclusion} will conclude our work.

\section{A brief history of blockchain innovations \label{lab:sec:history}}
Bitcoin is said to be the first blockchain application and is therefore considered the technology which has coined the term blockchain. The technology of Bitcoin is based on a whitepaper~\cite{nakamoto_bitcoin:_2008} published in 2008 by a person (or group) acting under the name Nakamoto (which is believed to be a pseudonym \cite{tasatanattakool_blockchain:_2018,peck_bitcoin:_2012}). It is based both on cryptographic principles (like \cite{back_hashcash} and \cite{merkle_protocols_1980}) and ideas concerning decentralized currencies (like \cite{szabo_bit_2008} and \cite{dai_b-money_1998}) that were published before. 

Nakamoto designed a digital currency system in which the coins are based on digital signatures. Doing so, they insured the security and integrity of coin transfers by using established cryptographic methods. In contrast to previous digital currencies like Szabos \textit{Bit gold} \cite{szabo_bit_2008} or Dais \textit{b-money} \cite{dai_b-money_1998} Nakamoto further solved the so-called double-spending problem. The double-spending problem describes the fact that digital goods might be duplicated easily. In contrast to physical money which can only be spent once (and is not in your possession thereafter), it is hard to assure that digital assets are only passed over to a single person or to make sure that such a multiple spending is at least noticed by others.    

\subsection{Technical implementation of the Bitcoin blockchain system}
Nakamoto solved the double-spending problem of previous decentralized currencies by defining that only the first attempt to spend a certain coin should be seen as valid. Later attempts should be rejected by the network. To make this possible, every member of the network has to be able to check the validity of any request to spend coins and to figure out if there have been previous requests of spending the same coins.

Coin transfers between owners is handled in transactions which contain information about the sender, the receiver, the amount of coins to be send and a proof that the sender is entitled to request the transaction (i.e, a proof that they are the owner of the coins they want to send). Proving the ownership of a certain coin is done using cryptographic concepts like digital signatures. Instead of relying solely on digital signatures to prove ownership, Nakamoto decided to introduce a script language that is used for this task. The decision came from considerations about enabling the modelling of complex requirements that need to be fulfilled to spend a certain coin. For example, it is possible to create a coin which can only be spent after a certain time period \cite{franco_understanding_2015}.  

Having transactions which can be validated with regard to the ownership of the coins is not enough to prevent double-spending attempts. To enable this, Nakamoto has decided that transactions are not broadcasted separately but in groups, the so-called blocks. %
Blocks are ordered and linked as a chain with each block pointing to its direct ancestor. By doing this, the order of the processed transactions can be captured. Grouping the transactions in blocks is still insufficient to solve the double-spending problem: a malicious member of the network could try to manipulate the blocks and leave out certain transactions. To prevent this, the blocks are protected against manipulation by a procedure Nakamoto called "proof-of-work". Proof-of-work is a cryptographic method introduced by Back \cite{back_hashcash} to prevent e-mail spam. The core idea is to include a proof that a certain amount of work has been done in order to demonstrate a certain commitment for the message being sent. In reality, this is often implemented by including a hash value that fulfills certain requirements (e.g. to start with a predefined number of zeros). Since hash functions are almost impossible to reverse or predict (i.e., finding the input for a given hash), the sender of the message has to try a lot of different input values (e.g. the message and a random value) to find one that leads to a hash value matching the requirements. Nakamoto used the same mechanism to secure the integrity of blocks. The hash value of a valid block has to start with a predefined number of zeros. If a block is changed, the hash value would change as well and, in most cases, not fulfill the requirements anymore. This enables the detection of illegal modifications \cite{antonopoulos_mastering_2015,franco_understanding_2015,decker_information_2013}.   

Besides the technical prevention of illegal modifications of issued blocks, the concept of Nakamotos blockchain depends on game theoretical elements as well \cite{finestone_game_2017}: Nakamoto decided that (at least in the reference implementation of Bitcoin) clients should accept the blockchain with the highest accumulated proof-of-work difficulty as valid \cite[p. 200ff.]{antonopoulos_mastering_2015}. Combined with the fact that the creation of new blocks requires a huge amount of computational power, this led to a system discouraging fraudulent behavior and promoting honesty \cite{lewenberg_bitcoin_2015,badertscher_but_2018,finestone_game_2017}. %
Even though there have been some problems with this assumption (see e.g. \cite{eyal_majority_2018,hern_2014,hornyak_2014,gervais_tampering_2015}), as far as we know, there has been no successful attack targeting this specific attack vector (cf. \cite{lee_2017} for successful ones). %
In the last years it has been shown that there are various problems with the concept presented by Nakamoto. Also, multiple attempts have been made to solve them. Some solutions are proposed as so-called Bitcoin Improvement Proposal (BIP)\footnote{\url{https://github.com/bitcoin/bips} (visited on 2018-08-09)}, others as completely new blockchain networks (e.g. Ethereum \cite{antonopoulos_mastering_2018}).

\subsection{Properties of blockchain technology}
According to Nakamoto the goal of Bitcoin was to create an electronic payment system that does not depend on trusted third parties and enables any two parties to transact directly with each other \cite{nakamoto_bitcoin:_2008}. Chatterjee \cite{chatterjee_overview_2017} states that there are five properties characterizing blockchain technology:

\begin{itemize}
	\item \underline{Immutable}: It should be practically impossible to modify a block.
	\item \underline{Irreversible}: It should not be possible to reverse a transaction once it was processed by the network participants.
	\item \underline{Distributed}: Everybody should be able to participate in the network and be able to validate and process transactions.
	\item \underline{No Centralized Authority}: There should not be a centralized organization the network depends on.
	\item \underline{Resilient}: The system should be able to handle faulty messages without opening up opportunities for fraud (also see \cite{conte_de_leon_blockchain:_2017}).
\end{itemize}

There are further properties mentioned in the literature which are paraphrasing those listed above. There are also very generic properties which can be used to describe other non-Bitcoin \textit{Decentralized Consensus Technology} (see Section~\ref{lab:sec:DCS}) as well. Furthermore, there are characteristics used to describe blockchain technology that we believe are applied wrongly. To make it more clear, we give an example for each category:

\begin{itemize}
	\item \textit{Properties paraphrasing the above ones} -- 
One prominent example for properties meaning the same but being described by different words is the \textit{distribution} of the system. As it was stated before, authors like Chatterjee \cite{chatterjee_overview_2017} or Nakamoto \cite{nakamoto_bitcoin:_2008} themselves see the fact that the system does not depend on a central service and that everybody is enabled to participate as core factors of the concept. In the literature \textit{"distributed"} and \textit{"decentralized"} are used synonymously in order to describe blockchain technology, ignoring the semantic differences between the two words \cite{xu_taxonomy_2017,bodrova_what_2017,walch_path_2017}. Section~\ref{lab:sec:decentralized:vs:distributed} will elaborate on the differences in more detail. Independent from the question which of both words is the correct one to characterize the technology, it serves as a good example to demonstrate the fact that different authors use different words to describe the same properties of blockchain concepts.

\item \textit{Too generic properties} --
Some properties used to describe blockchain-like systems are very generic. Conte de Leon states that being \textit{digital} is a fundamental concept of blockchain technology \cite{conte_de_leon_blockchain:_2017}. Without questioning this fact, it is clear that this property is too generic to describe and differentiate blockchain networks. 

Generic properties are not necessarily bad but they still need to define a boundary in order to be useful. For example there might be a superior category embracing both blockchain and non block- or chain-based \textit{Decentralized Consensus Technology}; independently of the internal functionality of such systems, being digital would also be a basic property.  %
Another example of these kind of properties is that blockchain technologies are said to be \textit{"trustless"}. It means that they do not rely on a (central) trusted third party \cite{cachin_blockchain_2017}. Some definitions even go one step further and declare that this trustlessness means that there has been a shift from trust in institutions or persons towards trust in algorithms and mathematical concepts (cf. \cite{conte_de_leon_blockchain:_2017,lustig_algorithmic_2015,grier_all_2014}). But similar to the digitality property, this also applies to a more generic term and not only blockchain technology.

\item \textit{Wrongly applied characteristics} --
In literature sometimes properties used to describe blockchain technology are applied wrongly. One example for an alleged property of blockchain technology is the \textit{anonymity} of blockchain users. The misconception that blockchain technology, respectively Bitcoin, is anonymous might go back to the fact that Nakamoto discusses transaction privacy in their white paper \cite{nakamoto_bitcoin:_2008}. To be more precise, they compare the privacy of transaction in their presented system to the traditional money systems. In that comparison Bitcoin is better since there is no trusted third party keeping records of all transactions but only hashed public keys as indicators on how much money belongs to a certain Bitcoin address. This might have led to the erroneous belief that Bitcoin is anonymous and could therefore be used for criminal reasons (see \cite{boehm_bitcoin:_2014,foley_sex_2018,chohan_cryptocurrency_2017}). Nevertheless, various studies (like \cite{bohr_who_2014,altshuler_analysis_2013,van_saberhagen_cryptonote_2013}) have shown that this assumption is not correct and Bitcoin can only be described as pseudonymous. More information on the anonymity of blockchain technology, and on the fact that only some specific blockchain systems offer their users anonymity, is given in Section~\ref{lab:sec:privacy}.          
\end{itemize}   

\subsection{Problem categories of Bitcoin technology \label{lab:sec:bitcoin:problems}}
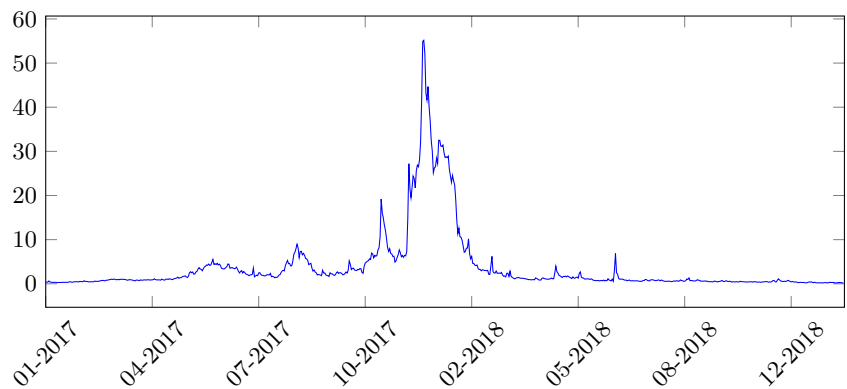
\begin{figure}
	\centering
	\begin{tikzpicture}
	\begin{axis}[
			date coordinates in=x,
			date ZERO=2017-01-01,
			xmin=2017-01-01,
			xmax=2019-01-21
			xtick distance=30,
			ytick distance=10,
			scaled y ticks=false,
			y tick label style={/pgf/number format/fixed},
			width=\columnwidth,
			height=\axisdefaultheight*0.75,
			xticklabel style={
				rotate=45,
			},
			xticklabel={\month-\year}
		]
		\addplot [blue, mark=none] table[x=date, y=tx_fee_usd] {tx_fee.dat};
	\end{axis}
	\end{tikzpicture}
	\caption{Average Bitcoin transaction fee in USD according to \url{https://bitinfocharts.com/comparison/bitcoin-transactionfees.html} (visited 2019-01-21) \label{fig:blockchain:transaction:fee}}
\end{figure}

Since the release of Bitcoin, several problems of the system have been shown. Some came up from practical considerations and some only from scientific examinations. Lin and Liao \cite{lin_survey_2017} identified several problems threatening the security and usability of the system: attacks focusing problems of the protocol, forks of the network, confirmation time of transactions, regulatory problems, scalability issues and integrated costs. %
We will now briefly discuss the categories \textit{attacks}, \textit{confirmation time and transaction costs} and \textit{scalability}.
\begin{itemize}
	\item \textit{Attacks} -- During the last years several attacks have been presented that could be used to publish fraudulent transactions, to double-spend coins or to trick other participants into believing that a certain transaction has been processed even though it has been rejected by the network. Examples for these kind of attacks are the so-called "Finney Attack" \cite{karame_two_2012}, "Race Attack" \cite{franco_understanding_2015}, "Eclipse Attacks" \cite{heilman_eclipse_2015} or "50\% + 1 Attack" \cite{eyal_majority_2018}. The first three focus on other users of the network and try to trick them into thinking that they successfully received their coins. Eclipse attacks try to achieve this by isolating the victim of the attack from the other members of the network. If this is possible, the attacker can fake the acceptance of their transactions so that the victim believes that the transaction has been processed successfully while it was actually rejected. The 50\% + 1 attack describes an attack vector in which the attacker controls the majority of the network's computational power and therefore is able to rewrite the transaction history. If this is possible, they could undo certain transactions and spend these coins again. Despite the name, recent studies have shown that it is not necessary to gain more than 50 percent of the total network computation power but that one third is sufficient~\cite{eyal_majority_2018}. Franco \cite{franco_understanding_2015} describes an attack called "transaction spamming" which could be seen as a form of denial-of-service attack. The attacker tries to flood the network (or certain members) with a huge amount of (useless) transactions so that they are not able to process the legitimate ones in time anymore.  
	
	\item \textit{Confirmation time and transaction costs} -- Other problems that prevent Bitcoin from being used for everyday services are the long confirmation time of transactions and the integrated costs of new transactions. The time until transactions become confirmed results from design decision by Nakamoto that a new block should be created roughly every ten minutes \cite{nakamoto_bitcoin:_2008}. This results in a confirmation time of a new transaction of roughly 60 minutes as the current best practice emerged that a transaction is valid if it is included in a block with more than six successors (at that point the probability that an attacker can modify enough blocks quickly enough to erase this transaction becomes acceptably small \cite{sompolinsky_secure_2015}). This long period until a transaction is securely included in the blockchain might complicate the development of certain use-cases and hinders the everyday usage of the technology. On the other hand, authors like Antonopoulos \cite[p. 247]{antonopoulos_mastering_2017} argue it is not necessary to wait until a certain transaction has been included irreversibly, only until the risk of fraud is acceptably small. 
	
	Another problem preventing Bitcoin from being used in daily life are the potentially high transaction costs. In general, though not strictly required, it is common to pay a small transaction fee besides the actual payment. This is done both to incentivize the other members of the network to process the own transaction and to prevent misuse of the network \cite{buterin_ethereum_2014}. One problem of this concept is the fluctuation and actual amount of the transaction fee that has to be paid. Figure~\ref{fig:blockchain:transaction:fee} illustrates this issue. From mid to end 2018 it shows an average transaction fee of about half a US-Dollar. But on the other hand the transaction fee peaked at nearly 55 USD in December 2017. Such volatility in the costs of a single transaction might prevent people from using Bitcoin for everyday services.
	
	\item \textit{Scalability} -- Scalability in the context of blockchain technology covers a wide range of different problems with the following three as the most threatening ones: Energy consumption, size of the blocks and total size of the blockchain. Energy consumption is an issue due to the concept of basing the security of the network on computational expensive routines, where a lot of energy is required for that task. De Vries examines this fact and concludes that the Bitcoin network consumes nearly as much electric energy as Ireland in 2018. There are calculations stating that the energy consumption will grow further extensively \cite{de_vries_bitcoins_2018}. The size of a single block is important for scalability considerations because it limits how many transactions can be included in one block and therefore processed in a certain time. In literature there is often a comparison between the number of transactions per second of Bitcoin and the amount of transactions in the same time frame of a traditional payment provider. The studies often compare Bitcoin's 7 transactions per second \cite{vukolic_quest_2015} with traditional payment providers like VISA which processes up to 56.000 transactions in the same time \cite{visa_inc._visa_nodate}. This comparison is not entirely correct anymore since the size limitation of a single block was increased in the Bitcoin system in August 2017 \cite{hrones_segwit_2017,redman_segregated_2017}. Instead of being a fixed size of 1 MB it was converted to a unit based system and increased to 4,000,000 units. Even when this would mean that more transactions could be processed in the same time as before, it is still no match for the traditional payment providers. Ehmke et al. \cite{ehmke_proof--property:_2018} point out that the growing size of the complete blockchain poses another problem considering new participants of the network. Since new members of the network need to download the whole blockchain before they can start to participate in the process of creating new blocks, an ever growing blockchain poses a challenge. If the blockchain should be used in the context of Internet-of-Things it might even prevent devices, respectively members, from participating at all because they might not have the storage to download the blockchain initially.
\end{itemize}

There are two general ways how the community handles changes to improve the technology and mitigate problems: The first one is the proposition of a so-called Bitcoin Improvement Proposal (BIP). These BIPs are a semi-formal way to propose changes in the network protocol (or the implementation) and follow a process also described in form of a BIP\footnote{ \url{https://github.com/bitcoin/bips/blob/master/bip-0001.mediawiki} (visited on 2018-08-16)}. The idea of BIPs is that new ideas are discussed in discussion groups growing in size until they reach the whole community. At that point the network members are able to vote in favor (or against) such an improvement and can therefore activate or reject it (see BIP 9\footnote{ \url{https://github.com/bitcoin/bips/blob/master/bip-0009.mediawiki} (visited on 2018-08-16)} for more information). If a BIP was accepted and implemented in the core code, it might lead to a fork of the blockchain. This can happen in situations where mining nodes do not accept the new software version or simply have not update their software yet. The term fork describes an event in which the blockchain is split into two or more separated chains with the same ancestor blocks. In reality there is a distinction between soft-forks and hard-forks. The difference between these two is the compatibility with blocks created by clients that do not follow the new protocol. Soft-forks are a type of fork in which the old clients can still process blocks produced by the new ones (e.g. when a certain rule has become more strict than before). Hard-forks means full incompatibility between the new and the old protocol version \cite{lin_survey_2017}. 

Another possibility to improve or modify the protocol is to create completely new systems. Due to the origin of blockchain applications (namely the usage as a currency) these new systems are often called "altcoins" or "altchains" \cite{antonopoulos_mastering_2015}. These terms describe both completely new networks or meta coin platforms which operate on top of existing blockchains but add extra features by inserting additional data into blocks or transactions. Even though some approaches try to enable the interoperability of different altcoins (see e.g. \cite{back_enabling_2014}), currently different chains are mostly unable to interact with each other \cite{underwood_blockchain_2016}.

\section{Evolution of the blockchain idea -- \newline The current state of the art \label{lab:sec:research}}
In the previous section we discussed that there are multiple properties a blockchain system is said to have. In this section an overview of the current state of the art, respectively the evolution of the blockchain concept in the last few years, is given. Each aspect that will be presented should demonstrate the modification of certain properties of the blockchain concept. %
There are approaches that change or remove properties that Nakamoto \cite{nakamoto_bitcoin:_2008} deemed essential but still make sense for certain use cases. Therefore a more general term describing technologies following the basic ideas of Nakamoto but not all of its concepts is required. %
 Nonetheless, the following survey shows the diversity of the current research in this topic and will show that a lot of the properties of Nakamotos system are being questioned. By examining the different modifications and improvements of the original idea by Nakamoto it will become apparent that there are a lot of attributes and properties the new approaches have or lack, which have not been discussed consciously but have been implicitly seen as granted.

\subsection{Privacy -- Between anonymity and pseudonymity \label{lab:sec:privacy}}
One of the properties that has been misinterpreted very often is the privacy of blockchain technology. Since its invention, Bitcoins decentralized character and its seeming anonymity \cite{gao_two_2016} led to a wide usage for criminal or illegal purposes \cite{van_wegberg_bitcoin_2018,yelowitz_characteristics_2015,bohr_who_2014,Kunnapas2016}. Nevertheless, studies like \cite{altshuler_analysis_2013} showed that the Bitcoin system is not anonymous but could be seen as pseudonymous.

Since Bitcoin is often used by people that do not want their financial transactions to be traceable (see \cite{bohr_who_2014,yelowitz_characteristics_2015}) different improvements (e.g. Zerocoin  \cite{miers_zerocoin:_2013,androulaki_hiding_2014}) have been made that also lead to completely new cryptocurrencies focusing on anonymity as a primary goal (see e.g. Monero \cite{kumar_traceability_2017,li_practical_2017}). 

Besides completely new and privacy-centered coins, there are other techniques such as the so-called \textit{mixing services} trying to anonymize payments by creating huge transactions with a lot of ingoing and outgoing addresses. The goal of these services is to complicate the analysis of the transactions and therefore to hide the sender and receiver of payments. This is done by mixing the actual payments with a lot of other, completely independent payments while it is only guaranteed that the receiver gets the agreed amount and not that it is received from a certain address (cf. \cite{chohan_cryptocurrency_2017,bonneau_mixcoin:_2014,ruffing_coinshuffle:_2014}).

Van Wegberg et al. \cite{van_wegberg_bitcoin_2018} point out that one of the problems of the anonymity of cryptocurrencies is that you often have to convert them back to a traditional (fiat) currency such as USD and EUR to be able to use them in everyday scenarios. Furthermore, the security of the system, independently of the security of its underlying protocol, often depends on the number of active users. Considering, for example, that Ripple is only maintained by very few servers (according to \cite{armknecht_ripple:_2015} the main network only consists of seven validating servers), the anonymity of the users can easily be compromised  without the usage of costly analysis like \cite{altshuler_analysis_2013}. Gervais et al. \cite{gervais_is_2014} point out that this critique is also valid for Bitcoin itself.

\subsection{Incentive for participation}
There is an ongoing discussion whether coins, respectively a native, inherent currency, is necessary for blockchains or if they could work without them. Blog posts like \cite{aru_can_2018} ask whether blockchain technology might could go on even when the inherent cryptocurrencies fail. Kravchenko \cite{kravchenko_does_2016} points out that there are different kinds of coins and tokens that could be used in blockchain environments. On the one hand there are inherent coins used in the protocol of the underlying blockchain and on the other hand there are tokens built on top of it (that might also be used in \textit{Initial Coin Offering} (ICOs)). The need for these two kind of coins, respectively tokens, has to be answered separately. More important is the discussion whether the inherent coins are needed for the working of the currency. There are different answers to that question. Cadigan \cite{silverstein_blockchain_nodate} states that there are different working blockchain projects without having the concept of inherent currencies. Private blockchains like R3's Corda or IBM's Fabric demonstrate that it is possible to construct systems without the need for inherent currencies \cite{perez_does_2017}. On the other hand these systems modify one important concept of previous blockchain systems, namely the openness for all people to participate (see Section~\ref{lab:sec:blockchain:types} for more details). 

Considering only public permissionless blockchains, in which everybody is allowed to participate, the reasons for an inherent currency were already given in Nakamoto's original paper: They state that the inherent currency helps to ensure that the participants stay honest \cite{nakamoto_bitcoin:_2008}. They justify this claim through the idea that, to create a stable system, staying honest has to be more profitable than trying to cheat. Therefore, a (real-world) incentive is needed to reward honest and punish dishonest participants \cite{carlsten_instability_2016,franco_understanding_2015}. Lewenberg et al. \cite{lewenberg_bitcoin_2015} and Badertscher et al. \cite{badertscher_but_2018} further support this claim by providing a game theoretical explanation for the stability of the system that relies on the concept of an external incentive for participants (like receiving coins which are valued by the participants and therefore serve as a currency).

Even though coins and tokens in blockchain systems are often called currencies, it is important to point out that in a legal definition they are not seen as such in most countries~\cite{Kunnapas2016}. Instead they are sometimes seen as commodities or digital goods. Moreover there is an ongoing discussion about regulations and whether some coins or tokens qualify as a security\footnote{cf. the SEC's statement which explains why e.g. Ethereum is not deemed as a security: \url{https://www.sec.gov/news/speech/speech-hinman-061418} or the Swiss FINMA regarding ICOs: \url{https://www.finma.ch/en/\~/media/finma/dokumente/dokumentencenter/8news/medienmitteilungen/20180216-mm-ico-wegleitung.pdf?la=en} (both visited on 2018-10-29)}. Independent of the classification of these inherent currencies, the current (tax) law has to be considered when coins and tokens are owned or used \cite{Kunnapas2016}. 

\subsection{Irreversibility and Immutability}
Properties, blockchain technology is often attributed for, are its irreversibility and  immutability. Irreversibility means that it is impossible to manipulate already issued blocks and is therefore the base for the immutability property of blockchains, which prevents a double-spending of coins \cite{chatterjee_overview_2017}. Newer studies demonstrate that these properties are not an intrinsic property of the technology, although they are often misunderstood as such. Conte de Leon et al. \cite{conte_de_leon_blockchain:_2017} point out that the blockchain is neither irreversible nor immutable. The probability that a certain already included transaction gets modified or removed, decreases over time so that it becomes unlikely, although it is not completely impossible \cite{conte_de_leon_blockchain:_2017}. 

One real-world example demonstrating the absence of both the irreversibility and immutability of blockchains is the Ethereum blockchain. In 2016 a bug in a smart contract deployed on the Ethereum blockchain (the concept of smart contracts will be described in more detail in Section~\ref{lab:sec:smart:contract}) led to a loss of coins worth nearly 60,000,000 USD (the "DAO-bug" \cite{atzei_survey_2017}). Since the loss was caused by a bug, respectively its active exploitation (and was considered a hack therefore), there has been a lot of discussions how to recover the coins \cite{noauthor_what_nodate}. In the end the majority of the network modified an old block and restarted the blockchain from that point on. Not all participants appreciated that solution which caused the network to split up into two parties, one using the "corrected" blockchain (Ethereum, \textit{"ETH"}) and one following the blockchain accepting the bug (Ethereum Classic, \textit{"ETC"}) \cite{kiffer_stick_2017}. This incident and the caused hard-fork demonstrate that blocks in the blockchain are not immutable at all. If enough members agree to modify a block or reverse certain transactions, they are able to do so.

Apart from bugs in the code of blockchains there might be other reasons why it could be necessary to be able to modify previous blocks: Matzutt et al. \cite{matzutt_quantitative_2018} published a study showing that the Bitcoin blockchain contains data that might be illegal to own and to share with others (e.g. child abuse imagery). Since every participant of the network shares the stored blocks with other (new) participants, the members of the Bitcoin blockchain distribute these illegal images. Strictly speaking, the fact that these kinds of images are included in the blockchain makes it illegal in some countries to participate as a miner node \cite{matzutt_quantitative_2018}.

At the time of writing, there are very few options on how these kind of data could be erased from the blockchain. One possibility is forcing a hard-fork of the blockchain (as it happened in the Ethereum network after the DAO-bug) and starting from a previous point in time again. This would mean that a lot of transaction data would be lost and payments therefore reverted. Since this would lead to a possible double-spending of the coins in newer blocks, this would jeopardize the whole concept of Nakamotos blockchain idea: preventing double-spending would not be possible anymore. Furthermore, at the moment there is no mechanism preventing the insertion of illegal data. In the last years there has been some research on the question of how a blockchain could be implemented offering the modification of previous blocks (see Ateniese et al. \cite{ateniese_redactable_2017} and Puddu et al. \cite{puddu_chain:_2017}). %
Another possible solution that might minimize the risks of these problems is the idea of processing transactions off-chain (see Section~\ref{lab:sec:off:chain} for more information on that topic) or enabling mechanisms in which not all data of the blockchain have to be cached to be able to participate as a full member in the network \cite{ehmke_proof--property:_2018}.

For developers of smart contracts the irreversibility and immutability poses a problem for debugging and upgrading their software. As we have seen in the example before, a bug in a smart contract can usually only be fixed by deploying a new version of it. Changes to an existing contract are not possible although there are some design patterns that support dynamic dispatching of function calls, e.g., the proxy\footnote{\url{https://blog.zeppelinos.org/proxy-patterns/} (visited on 2018-10-24)} or proxy delegate pattern\footnote{\url{https://fravoll.github.io/solidity-patterns/proxy_delegate.html} (visited on 2018-10-24)}, the \texttt{DELEGATECALL} opcode in Ethereum\footnote{\url{https://solidity.readthedocs.io/en/latest/introduction-to-smart-contracts.html\#delegatecall-callcode-and-libraries} (visited on 2018-10-24)} or registry contracts.

\subsection{Trustlessness -- Reaching consensus between participants \label{lab:sec:blockchain:types}}
Trust is one of the core elements of blockchain technology and being able to maintain consensus about a shared ledger without the need to trust a central party or any participant is one of the main features blockchain technology enables. It is sometimes said that blockchain based systems represent a shift from trust in institutions (like banks and governments) to trust in cryptographic principles \cite{conte_de_leon_blockchain:_2017,lustig_algorithmic_2015}. Nakamoto describes the goal of their system as allowing two parties to transact without the need for a trusted third party \cite{nakamoto_bitcoin:_2008}. 

A central question in this context is which cryptographic principles, respectively methods how the network agrees on a commonly shared state, are used to replace the formerly trusted third parties. Nakamoto builds upon the concepts of Back \cite{back_hashcash}, Dwork and Naor \cite{dwork_pricing_1993} and proposes to use a procedure they call "\textit{proof-of-work}" (PoW). This makes the creation of blocks a resource- and time-consuming process and they define that the chain of blocks with the highest accumulated proof-of-work is seen as valid. As mentioned in Section~\ref{lab:sec:bitcoin:problems} this procedure is often criticized due to its high energy consumption.

Multiple alternatives with different benefits and disadvantages are being discussed in literature. Some of them will be discussed in the following:
\begin{itemize}
	\item \textit{Practical Byzantine Fault Tolerance} (PBFT) is a technique by Castro and Liskov \cite{castro_practical_1999} that enables distributed members of a network to efficiently agree on a common state. One disadvantage of this procedure is that the participants of the network have to be known in advance \cite{vukolic_quest_2015}. It is therefore not possible to be used in systems such as Bitcoin that want to enable anyone to join and leave the network at any time.
	\item \textit{Proof-of-Stake} (PoS) modifies the initial proof-of-work idea by making it easier for participants who already own coins to create a new block (the more coins somebody owns, the easier it is to create a new block) \cite{bentov_cryptocurrencies_2016,tschorsch_bitcoin_2016}. The core idea of the concept is that participants who already own coins are willing to contribute work to stabilize the network because the value of their coins depends on it. Attacking the network would destabilize it and therefore jeopardize the value of their own coins as well. Giving participants who own coins, and therefore are interested in a stable network, more responsibilities and options to contribute to that goal, seems to be conceived as a good idea. Nonetheless, some researchers indicate there might be attack vectors preventing a real-world usage \cite{poelstra_distributed_2014}.
	\item \textit{Proof-of-Activity} tries to mitigate the shortcomings of proof-of-work and proof-of-stake algorithms by proposing a schema in which multiple users have to work together to create a new block \cite{bentov_proof_2014}. It could be seen as a combination of the voting idea of PBFT-protocols with the idea of the proof-of-work concepts that the right to create blocks should require dedicated work. Unfortunately, as far as we know, these kinds of methods are not widely adapted and therefore also lack a deep practical evaluation.
	\item \textit{Proof-of-Elapsed-Time} is an example of a schema that relies on trusted computing hardware and uses this to make the creation of blocks verifiable time-consuming \cite{chen_security_2017}. One example of this technique is the Hyperledger Sawtooth project \cite{olson_sawtooth:_2018}. Unfortunately, this technique requires the users to trust in specific hardware vendors, which could be seen as a contradiction to Nakamoto's idea to trust in algorithms instead of specific institutions.
	\item \textit{Other approaches}: Apart from the previous concepts, there are also various other attempts to replace the proof-of-work technique with ones that are more suitable in certain situations. Examples for this are "more useful" proof-of-work algorithms like the Primecoin system in which chains of prime numbers are searched \cite{king_primecoin:_2013}, \textit{proof-of-burn} schemata \cite[p. 236f.]{franco_understanding_2015}, \textit{proof-of-capacity} \cite{patterson_alternatives_2016} (e.g. used in Storj \cite{wilkinson_storj_2016}) or hybrid approaches as combinations that try to mitigate the disadvantages of the single concepts (cf. \cite[p. 173f.]{franco_understanding_2015}, \cite{bradbury_blocks_2015}).
\end{itemize}        

As previously mentioned the different concepts have various benefits, problems and restrictions. Some researches even state that reaching consensus in an asynchronous system in which a single entity might fail, is impossible altogether \cite{fischer_impossibility_1985,ben-or_completeness_1988}. 

One general distinction between the types of system is the ability for new participants to join the network. The importance of this difference even led to a distinction in terminology for blockchain systems steering more towards a closed or open direction. %
There are three broad categories of blockchain systems: Private blockchains, consortium blockchains and public blockchains \cite{wust_you_2017,xu_taxonomy_2017,goranovic_blockchain_2017}. %
Each category offers a different degree of trustlessness and level of security. %
In \textit{private blockchains} write permissions are restricted to a single entity or company who is responsible for adding data. Therefore, a distributed consensus mechanism is not required. The stored data can be visible to the public or its read access restricted. This blockchain variation is highly centralized but could offer easier auditability than common databases by using cryptographic elements. %
\textit{Consortium blockchains} employ a consensus process which is usually controlled by a predefined set of participants (i.e.~the consortium). Therefore, write permissions are partially decentralized and adding a block to the shared ledger requires the majority of the participants to confirm its validity. Reading the ledger can also be public or limited to the consortium itself. Consortium blockchains are sometimes called hybrid blockchains as they are located between private blockchains, with a single trusted entity, as one end of the spectrum and public blockchains, without the need for trust in any participant, on the other end. %
\textit{Public blockchains} allow anyone to read the current state and send valid transactions to be included in the shared ledger (i.e.~writing data to the blockchain). The consensus process is run by many participants who do not necessarily know or trust each other as anyone can participate and support the process. This results in a highly decentralized system.

\subsection{Operational purpose \label{lab:sec:chain:generations}}
Blockchain technology has been developed initially with the goal to be an alternative to the traditional money system and was extended over time to become the base of various use cases \cite{swan_blockchain:_2015}. In literature this evolution is separated into three different stages. Even though different names for the layers exist, they basically mean the same: Swan \cite{swan_blockchain:_2015} and Zhang et al. \cite{zhang_towards_2018} call them \textit{"Blockchain~1.0"}, \textit{"Blockchain~2.0"} and \textit{"Blockchain~3.0"}, whereas Bodrova \cite{bodrova_what_2017} and Johnston \cite{johnston_decentralizedapplications} call them \textit{"Type~I DApps"}, \textit{"Type~II DApps"} and \textit{"Type~III DApps"} (where DApp stands for \textit{Decentralized Applications}). They share the same understanding of the different layers: The first layer consists  only of cryptocurrencies (like Bitcoin) that use an own blockchain operated by the users of the corresponding currency. The second layer is based on this, but enhances it by providing smart contracts that could be used to transfer assets of any kinds between the members of the network. Applications of the second layer generally do not use an own blockchain but are part of an existing one in which they are integrated through issued tokens. The third layer describes all other applications that could be build on top of a blockchain but that are not focused on asset transactions.   

\paragraph*{Blockchain 1.0 -- Cryptocurrencies} As already stated Nakamoto designed their system as an alternative to the traditional financial system which relied on trust in different institutions. Instead the system should only rely on the soundness of mathematical concepts (respectively algorithms)~\cite{nakamoto_bitcoin:_2008}. A lot of subsequent altcoins followed this approach. The website coinmarketcap.com\footnote{\label{footnote:conmarketcap}\url{https://coinmarketcap.com/all/views/all/} (visited on 2018-10-05)} lists more than 2.000 different coins at the time of writing. Even though not all listed coins are solely designed as cryptocurrencies, but provide solutions for the other two layers of blockchain applications as well, some of the oldest and widely used currencies (like Litecoin, Dogecoin or Ripple) fall in this category \cite{swan_blockchain:_2015}.  

\paragraph*{Blockchain 2.0 -- Smart Contracts and Decentralized Code Execution\label{lab:sec:smart:contract}}
One of the design decisions of the Bitcoin protocol was the limitation of its script language (which is used to prove the ownership of a certain address and therefore of the coins associated with it) to being explicitly Turing-incomplete \cite{franco_understanding_2015}. The decision was justified by the need to find a way to prevent attackers from creating transactions with infinity loops that might lock the verifier of the transaction in never-ending calculations and prevent them from processing legitimate transactions \cite{antonopoulos_mastering_2017}. At the moment the Bitcoin reference implementation even further restricts the flexibility of the script language by accepting only certain predefined types of transactions \cite{franco_understanding_2015}.

In 2012 a proposal called "The Second Bitcoin Whitepaper" was published trying to enhance the flexibility of conditions that have to be fulfilled to validate a transaction \cite{willett_second_2012}. The proposal led to more general propositions about ways to embed small programs in the blockchain \cite{buterin_prehistory_2017,kokoris-kogias_omniledger_2018}. In 2013 the project Mastercoin published its whitepaper and proposed an alternative to Bitcoin that allows its users not only to trade assets but to share small programs with others. These programs would be executed by all participants of the network when triggered and could therefore change the global state of the network as it is replicated by the different participants. These initial ideas were the base for the development of Ethereum~\cite{wood_ethereum:_2014} which is the second most valuable blockchain at the moment (according to coinmarketcap.com$^{\ref{footnote:conmarketcap}}$). %
Currently there are projects trying to enable users of the original Bitcoin network to execute smart contracts as well. For example \textit{Rootstock/RSK} \cite{lerner_rsk_2015} and \textit{Counterparty} \cite{noauthor_protocol_2018} try to establish this feature through sidechains \cite{bocek_smart_2018} (see Section~\ref{lab:sec:off:chain} for more information).

One important aspect is the fact that smart contracts are only the foundation of the second layer of blockchain applications. Antonopoulos \cite{antonopoulos_mastering_2018} points out that decentralized applications consist of a (web) front-end, data storage, communication layers and the back-end software. Only the back-end (or sometimes even only parts of it \cite{wessling_how_2018}) are implemented as smart contracts. 

\paragraph*{Blockchain 3.0 -- Further applications}
The third layer of blockchains comprises all applications beyond financial transactions. Antonopoulos states that there are various use cases for blockchain technology and financial applications are only the first \cite{andreas_m._antonopoulos_introduction_2016}. Blockchain technology is often said to be for new digital businesses what TCP/IP has been for the internet: to serve as a technology base layer enabling new use cases \cite{iansiti_truth_2017}. The possibilities for use cases range from medical applications \cite{azaria_medrec:_2016,swan_blockchain_2015} over usages in state government \cite{bartolucci_sharvot:_2018,olnes_blockchain_2017} to usage in the context of Internet-of-Things (IoT) \cite{gries_using_2018,kshetri_can_2017}. There is ongoing research covering the topic on how to determine whether a certain business idea might benefit from using blockchain technology (see e.g. \cite{wessling_how_2018,wust_you_2017,lo_evaluating_2017,peck_blockchain_2017}).

Since there are very different business cases in which blockchains can be integrated, there are also various ways how these applications interact with the blockchain. Some use the blockchain e.g. to record the timestamps of data through inserting their hashes into an existing blockchain (see digital notary services like OriginStamp \cite{gipp_decentralized_2015} or Stampery \cite{de_pedro_crespo_stampery_2017}). Other projects try to establish a decentralized market using blockchain technology (see e.g. \cite{pop_blockchain_2018}). Both concepts have to face different challenges. The idea of using existing blockchains to store arbitrary data leads to several discussions about the technical possibilities (e.g. the introduction of the \textit{OP\_RETURN} operator for the Bitcoin validation script \cite{bartoletti_analysis_2017}) and about the general reasoning for storing data on the blockchain at all (e.g.~the Bitcoin Foundation stated that Bitcoin is not designed for this task \cite{bradbury_blocks_2015}). Furthermore, since the data included in the blockchain is distributed to all participants and cannot be erased easily, once it was included, serious legal concerns came up (see e.g. \cite{matzutt_quantitative_2018}). Using the blockchain for non-financial use cases often imposes another difficult problem: how can smart contracts that are deterministic and are not able to depend on data outside the blockchain interact with real-world data? One practical example for this kind of problem is a decentralized insurance policy, e.g. where an immediate compensation for delayed flights is implemented\footnote{see e.g. \url{https://fizzy.axa} (visited 2018-10-29) or P{\"u}ttgen and Kaulartz \cite{puttgen_versicherung_2017}}. The smart contract of the insurance should automatically activate the payment process in that case, but due to its nature is not able notice delays (or other real-world events). Services like oraclize.it \cite{noauthor_oraclize_2018} offer a solution to this problem by recording real-world events in the blockchain (where it is accessible by the smart contracts again), but the research is still ongoing \cite{swan_blockchain_2015}.

\subsection{Confirmation time and transaction costs \label{lab:sec:time:and:costs}}
From a business perspective, there are two major flaws with blockchain technology: On the one hand, sellers are required to wait for a given time until it is certain that a payment transaction cannot be reversed and on the other hand, publishing a transaction costs a certain fee that might outstrip the value of the actual payment \cite{bamert_have_2013}. In the last years a lot of research has been done on how to handle this and and how this influences the operational costs of the resulting software architecture (e.g. \cite{wessling_tactics_2019}). An easy solution to reduce the confirmation time would be to decrease the average time between blocks (in Bitcoin this is about ten minutes). Several altcoins and alternative blockchains have done so \cite{beck_blockchain_2016}. Unfortunately, if the blockchain is designed as a linear chain of blocks (as its name suggests) this might lead to problems since a smaller blocktime inevitably leads to a higher rate of forks in the blockchain (cf. \cite{decker_information_2013}), which in turn leads to a waste of energy since all concurrent blocks but the ones of the longest chain (respectively that one with the highest cumulated proof-of-work) are dismissed \cite[p. 247]{antonopoulos_mastering_2017}. %
Sompolinsky et al. \cite{sompolinsky_secure_2015} present a procedure for interpreting the blockchain as a directed graph and taking sibling blocks into consideration when determining which path of the blockchain is the valid one. Because of this, the computational power included in sibling blocks is not completely wasted anymore, which reduces the need to avoid forks. The (adapted) implementation of the proposed procedure enabled alternative blockchains like Ethereum to reduce their block time significantly (on average 12~seconds for Ethereum versus 10~minutes for Bitcoin) \cite{buterin_ethereum_2014}.

Other concepts follow this idea and try to take the "stale" blocks, which would be dismissed in Bitcoin-like concepts, into consideration when building the block structure. For example IOTA's data structure \textit{The Tangle} \cite{popov_tangle_2018} groups transactions in a directed acyclic graph (DAG) instead of a single linear chain. In this case the energy used to create the blocks not belonging to the main chain would not be wasted but used to further stabilize it. Another benefit of this procedure is that the transactions grouped in sibling blocks would not be ignored by the network but would belong to the replicated data structure known by all participants independently of the fact that they are not part of the main chain. \textit{Hashgraph} \cite{baird_methods_2017,baird_swirlds_2016} proposed to use an adapted consensus protocol in which the constantly forking of the chain is intended and the agreement of the other participants of the network to a certain network state can be proven mathematically instead of waiting for them to explicitly express it. This procedure leads to a system in which forks are nothing negative that should be avoided and is therefore able to issue blocks with a very small block time. To give a holistic picture it is necessary to add that, to our knowledge, \textit{Hashgraph} is not widely used yet (maybe due to the fact that it is patented \cite{baird_methods_2017}) and that \textit{The Tangle} was criticized a lot because it is difficult for researchers outside the IOTA company to independently evaluate the proposed technology (cf. \cite{peck_cryptographers_2018}).   

There are also projects trying to solve the confirmation time problem of Bitcoin without using a DAG data structure. Eyal et al.~\cite{eyal_bitcoin-ng:_2016} propose a concept that does not only use blocks but also so-called microblocks that are issued every 10~seconds (while the "full" blocks are still issued roughly every 10~minutes). The microblocks are issued by the creator of the last full block and do not require a proof-of-work. Therefore, the network is able to process more transactions in less time compared to a (Bitcoin-like) system in which every block is secured by such computations.

Apart from the confirmation time of transactions, the second problem that prevents the wide usage of blockchain technology is the required transaction fees. Nakamoto \cite{nakamoto_bitcoin:_2008} introduced them as an incentive for the miners to process transactions of other users. In an addition to the transaction fee there is also a block reward that is used to incentivize miners to participate in the Bitcoin network. Because Bitcoin is designed as a currency with a fixed maximum amount of coins, the block reward is halved roughly every four years \cite[p. 214]{antonopoulos_mastering_2017}. The halving of the mining reward might has some interesting consequences. While it is possible that the transaction fee could simply rise to compensate the miners for the decreasing block reward (even though studies from 2015 \cite{moser_trends_2015} and the actual transaction fee development of the last years (see Figure~\ref{fig:blockchain:transaction:fee}) seem to disprove this theory), there might also be other consequences like attacks on the network to validate only certain transactions with an especially high transaction fee (cf. \cite{carlsten_instability_2016}). There are research projects trying to solve these problems, e.g. the IOTA project states that there are no transaction fees in their network\footnote{see paragraph "Zero fees" at IOTAs Q\&A page (\url{https://iotasupport.com/whatisiota.shtml} (visited on 2018-10-16))}. Projects like Soluna \cite{belizaire_soluna_2018} try to use renewable energy to mine new blocks and might be able to do so economically without requiring huge transaction fees. Furthermore, there are concepts like off-chaining and sidechains trying to separate multiple small transactions from the main chain (using it only as a security reference) that might be able to offer solutions in which the transaction costs could be reduced significantly compared to a solution using the main blockchain for every transaction (see the following Section~\ref{lab:sec:off:chain} for more information).

\subsection{Scalability \label{lab:sec:maturing:scalability}}
Several problems of blockchain technology are summarized under the term \textit{scalability}. One is the amount of transactions the system can process in a certain time. Since Bitcoin as the origin of blockchain technology is seen as an alternative to the traditional payment processors, Bitcoins ability to process transactions is often compared to traditional credit card vendors. It is often stated that Bitcoin can process a maximum of 7~transactions per second \cite{gervais_tampering_2015} while Visa is said to be able to process a peak number of 56,000~transactions per second \cite{visa_inc._visa_nodate}. Independent of the question of how long it takes until a transaction is (practically) irreversibly included in the blockchain (see the previous Section~\ref{lab:sec:time:and:costs}) this is quite an important consideration since it is often sufficiently secure enough when a certain transaction was included in the blockchain at all (cf.~\cite[p. 25]{antonopoulos_mastering_2015}).

One solution for this problem is called "Segregated Witness" (SegWit) and was proposed in the Bitcoin Improvement Proposals BIP-141, BIP-143, BIP-144 and BIP-145\footnote{the BIPs can be found here: \url{https://github.com/bitcoin/bips} (visited on 2018-10-17)}. In addition to other improvements the core idea of the concept was to change how the size of transactions is counted. In fact, following the proposal, the locking script that is used to prove the ownership of a certain coin to be spent, will not be counted anymore when calculating the size of a transaction. This led to a mechanism which allows more transactions to be included in a block while maintaining the same blocksize as before (which is fixed at 1\,MB) \cite[p. 329ff.]{antonopoulos_mastering_2017}. For a proposal to additionally increase the maximum block size to 2\,MB no consensus between the participants could be reached \cite{vigna_bitcoin_2017}.

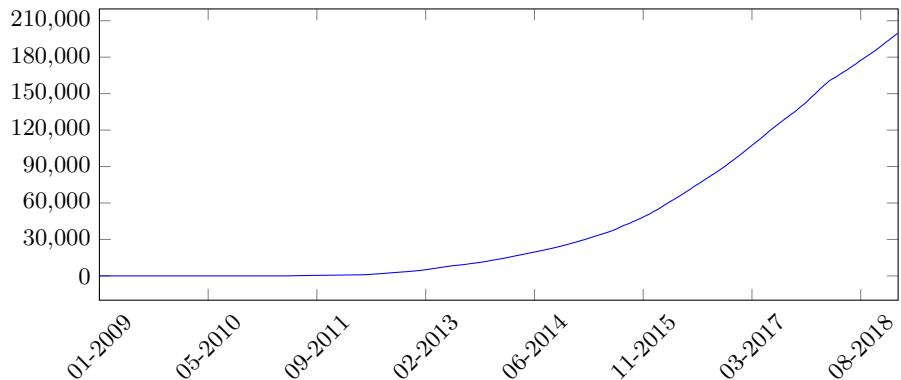
\begin{figure}
	\centering
	\begin{tikzpicture}
		\begin{axis}[
			date coordinates in=x,
			date ZERO=2009-01-01,
			xmin=2009-01-01,
			xmax=2019-01-21
			xtick distance=180,
			scaled y ticks=false,
			ytick distance=30000,
			y tick label style={/pgf/number format/fixed},
			width=\columnwidth,
			height=\axisdefaultheight*0.75,
			xticklabel style={
				rotate=45,
			},
			xticklabel={\month-\year}
			]
			\addplot [blue, mark=none] table[x=date, y=size] {blockchain_size.dat};
		\end{axis}
	\end{tikzpicture}
	\caption{Size of the Bitcoin blockchain in MB according to \url{https://www.blockchain.com/de/charts/blocks-size} (visited on 2019-01-21) \label{lab:fig:blockchain:size}}
\end{figure}

Another aspect of scalability is the way how new participants can be integrated into the network: How easily can the network scale by distributing it to more network members? Although this question is also influenced by the block size, the accumulated size of the whole blockchain is relevant here. The Bitcoin system expects each full member to validate incoming transactions which is only possible if they know about the current state (respectively the amount of coins all addresses possess) of the network. Since the current state of the network is not distributed directly, but has to be calculated by processing all accepted transactions of the network again, new participants have to download the whole blockchain \cite[p. 32ff]{antonopoulos_mastering_2015}. The Bitcoin blockchain contains of more than 180\,GB of data at the time of writing (Figure~\ref{lab:fig:blockchain:size} visualizes the growing of the Bitcoin blockchain over time). The huge amount of data that has to be downloaded by new participants might be a barrier for their involvement~\cite{ehmke_proof--property:_2018,zheng_overview_2017}. It is sometimes argued that new participants do not necessarily need the whole blockchain because there are lightweight verification protocols (cf. "Simplified Payment Verification" (SPV) in Nakamoto's initial paper~\cite{nakamoto_bitcoin:_2008}), but they rely on other participants to have downloaded the whole blockchain and do not enable its users to create new blocks. Solutions like \cite{ehmke_proof--property:_2018} or \cite{skudnov_mini-blockchain_2017} try to introduce approaches in which old transactions could be dismissed and do not have to be downloaded by new participants anymore. Therefore space and time requirements for new participants could be reduced, which in turn might encourage more people to actively participate in the verification and creation of blocks. 

Other approaches like sidechains and off-chaining (as explained in the following section) propose to reduce the number of operations required to be executed on the main chain and therefore would help to slow down the growing of the chain. Nevertheless, new participants would still need to download the previous data.

\subsection{Sidechains and Off-Chaining \label{lab:sec:off:chain}}

Scalability as outlined in the previous section poses several challenges for designing blockchain technologies and has to be kept in mind when creating decentralized applications. %
Buterin et al. summarize the challenges of blockchain technology in three fundamental properties\footnote{cf. \url{https://github.com/ethereum/wiki/wiki/Sharding-FAQs} (visited on 2018-10-22)}: Decentralization, scalability and security. %
The "scalability trilemma" (cf. \cite{zhang_towards_2018}) states that it is not possible for a blockchain system to fully support all three properties simultaneously but only two of them. There are three possible combinations: %
(1) A centralized system (e.g., a web server) offers high scalability and security by being able to add resources on demand (e.g., by adding more RAM to its virtual machine). %
In contrast, (2) a highly decentralized system offers high security by distributing its elements on many nodes, which comes with the drawback of low scalability. %
(3) A high degree of decentralization and high scalability can be achieved by simplifying the consensus process, which in turn results in less security. %
For example the current implementation of Ethereum focuses on decentralization and security, whereas scalability remains as an open issue. %
There are several proposed solutions which are currently under examination and development. %
Sidechains are built on top of a main blockchain (e.g. Ethereum) and can use a custom consensus algorithm with own public nodes (e.g. Plasma\cite{poon_plasma:_2017} or Loom Network DAppChains \cite{duffy_everything_2018}). This concept increases the performance, makes all transaction on the sidechain publicly visible and simultaneously secures the execution of transactions as the sidechain is connected to the main chain. 

In contrast to sidechains, "off-chain" interactions are executed independently from the main chain \cite{eberhardt_or_2017}. During the interaction between two parties, signed receipts of small transactions are exchanged and only the final result will be written to the main chain (in case both parties agree). %
Off-chaining concepts can be separated into three categories: %
\textit{(1)~Transactional off-chaining} aims at exchanging tokens (e.g. ERC-20 tokens in Raiden~\cite{noauthor_raiden_2018}) or coins (Bitcoins in Lightning~\cite{poon_bitcoin_2016}) between two parties via channels that are separated from the main chain. %
\textit{(2)~Storage off-chaining} aims at providing a means to store arbitrary data (e.g. text and binary files) usually to a decentralized storage network such as IPFS~\cite{benet_ipfs_2018}, Sia~\cite{vorick_sia:_2014} or Storj~\cite{wilkinson_storj_2016}. This is due to the fact that storage is very expensive on blockchains such as Ethereum and Bitcoin~\cite{rimba_comparing_2017}. %
\textit{(3)~Computational off-chaining} such as zero-knowledge proofs (see e.g. Zcash, zkSNARKs) can be used because complex computations are not only expensive to be executed, e.g. in a smart-contract, but can also be technically impossible due to the blocksize limit which allows only a certain amount of instructions per block.

\subsection{Centralized, Decentralized and Distributed Architectures \label{lab:sec:decentralized:vs:distributed}}
\begin{figure}
	\centering
	\includegraphics[width=0.9\columnwidth]{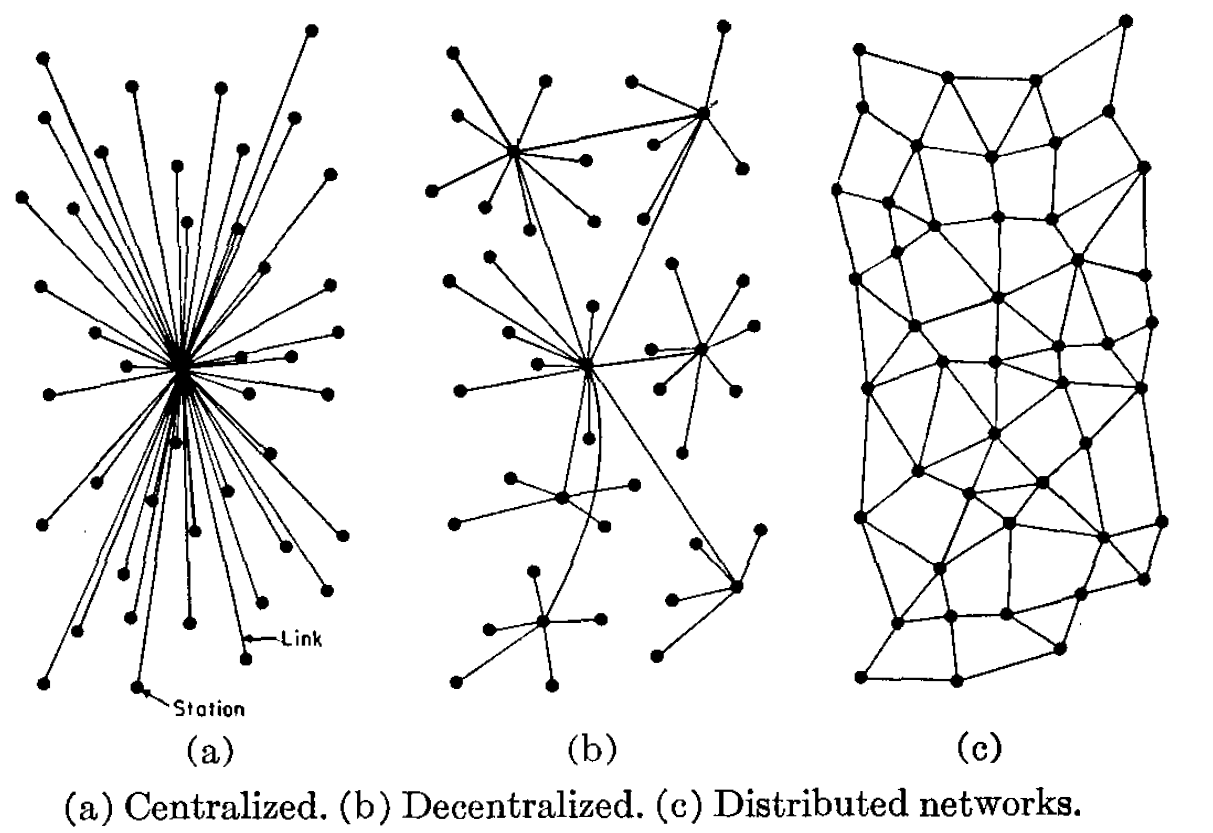}
	\caption{Distinction between centralized, decentralized and distributed systems according to Baran (Figure taken from \cite{baran_distributed_1964})\label{fig:decentralized:distributed}}
\end{figure}

In the Bitcoin paper Nakamoto \cite{nakamoto_bitcoin:_2008} states that their technology should enable two parties to transact directly with each other. Later, they use the term "peer-to-peer distributed timestamp server" \cite{nakamoto_bitcoin:_2008} to describe the proposed Bitcoin system as a solution to enable this. This wording is controversial because in literature the terms "\textit{decentralized}" and "\textit{distributed}" are often used synonymously to describe blockchain-like systems (see e.g. \cite{xu_taxonomy_2017,bodrova_what_2017,walch_path_2017}). Internet discussions (e.g. \cite{buterin_meaning_2017} or \cite{bodrova_what_2017}) often refer to an explanation of Baran \cite{baran_distributed_1964} to demonstrate the differences between these two terms. Figure~\ref{fig:decentralized:distributed} is taken from the original paper from Baran and visualizes their distinction: In a centralized system all nodes are connected to a single central node. In a decentralized system there are multiple nodes serving as bridges between the others. In a fully distributed system any node is connected with all the others via more than one route and therefore preventing the creation of essential nodes, which would split the network in case of an outage.

Independent from the distinction whether Bitcoin is \textit{distributed} or just \textit{decentralized}, it is sometimes questioned whether it is even not-centralized at all. Arguments against seeing the technology as not-centralized criticize the central role the developers of the reference implementation, mining-pool operators and the online wallet and exchange operators play \cite{beikverdi_trend_2015,gervais_is_2014,gencer_decentralization_2018}. Baran uses their distinction between the terms \textit{centralized}, \textit{decentralized} and \textit{distributed} mainly to discuss the ability of the network to handle failures of (important) parts of the network and its resilience against attacks. Both aspects have also be considered when discussing whether current blockchain systems are not-centralized at all. Event though there is ongoing research trying to develop consensus mechanisms that cannot be outsourced to mining pools \cite{ekblaw_bitcoin_2016} and that are not easily adaptable for specialized hardware \cite{zamanov_asic-resistant_2018,ren_bandwidth_2017}, it still has to be questioned whether the current networks can handle specific attacks while the consensus mechanism heavily rely on certain hardware and vendors (cf. \cite{wilmoth_bitmains_2018}). %
In an online article Buterin \cite{buterin_meaning_2017} picks up this discussion and argues that the distinction of Baran (see Section~\ref{fig:decentralized:distributed}) is misleading. Instead they propose to consider three dimensions of (de-)centralization: 
\begin{itemize}
	\item Architectural (de-)centralization -- How many physical computers is a system made up of?
	\item Political (de-)centralization -- How many individuals or organizations practically control these computers?
	\item Logical (de-)centralization -- Is the system, although distributed to multiple physical computers, representing \textit{one} logical unit?
\end{itemize}   

They argue that blockchain systems are often politically decentralized (since no single organization controls them) and architecturally decentralized (since the system is designed to be distributed to a lot of physically independent computers) but logically centralized (since it shares one logical state all participants agree on) \cite{buterin_meaning_2017}.
Nonetheless, their claim that systems like Ethereum are politically decentralized is often doubted: First, as already said, because of the central role of mining pools, exchanges and online wallet operators, second because of the important role of the reference implementation developers. Laurie \cite{laurie_decentralised_2011} argues that e.g. Bitcoin implemented a checkpoint concept in its blockchain to increase scalability and make the integration of transactions more deterministic (since it is not longer possible to modify the chain before the newest checkpoint, data in these blocks are not only unlikely to be modified anymore but it is nearly impossible to do so). The members of the network have to agree which block is considered a checkpoint. Laurie argues that this is done by a group of developers and not in a decentralized manner \cite{laurie_decentralised_2011}. To counter this critique, the Ethereum community implemented a decentralized procedure that defines checkpoint blocks \cite{buterin_casper_2017}. 

Other blockchain implementations like private or consortium chains (see Section~\ref{lab:sec:blockchain:types}) might lead to a different degree of (de-)centralization. Therefore, it is still an area of ongoing research and should be considered when developing blockchain applications (cf. \cite{wessling_how_2018}).  

\section{Decentralized Consensus Technology \label{lab:sec:DCS}}
The huge amount of projects and research, both from academia and industry, concerning blockchain technology shows that it is a topic of huge interest and rapid development. In the previous Section~\ref{lab:sec:research} we gave an overview about the wide range of research topics, the problems researchers are trying to solve and the benefits and disadvantages of those solutions. Furthermore, the overview shows that there is a quick evolution of the initial concept presented by Nakamoto. Several of the presented concepts modify or remove core concepts of Nakamotos approach but there are also several projects adding new abilities to the systems they develop that were not mentioned or envisioned in the initial paper of the technology. 

These variations of the concept of Nakamoto led to an interesting question: Is the term "blockchain" still appropriate to describe a technology which has modified core concepts of the blockchain idea? To give a more catchy example: Schiener, co-founder of IOTA, describes the IOTA concepts as being a "Blockchain without Blocks and the Chain" \cite{schiener_primer_2017}. Do we still consider such a system a blockchain or is another term needed to describe such a huge deviation from the initial concept?

We argue that the latter is the case. A proverb, Sokrates is often attributed for, states that the beginning of wisdom is the definition of terms. This might also be true for technical topics. We believe that a lot of the projects presented in the previous section are not modified versions of the initial Bitcoin-like concept but instead developed into independent technologies. To describe them accurately we propose to use the term "\textit{Decentralized Consensus Technology}" (DCT). Note that this term is not new: Chen et al.~\cite{chen_efficient_1992} already used the term \textit{Decentralized Consensus System} in 1992 within a slightly different context and Glaser et al.~\cite{glaser_beyond_2015} picked it up in context of blockchain systems. Unfortunately, to our knowledge, this terminology is not widely used in literature yet. 

\begin{figure}
	\centering
	\includegraphics[width=\columnwidth]{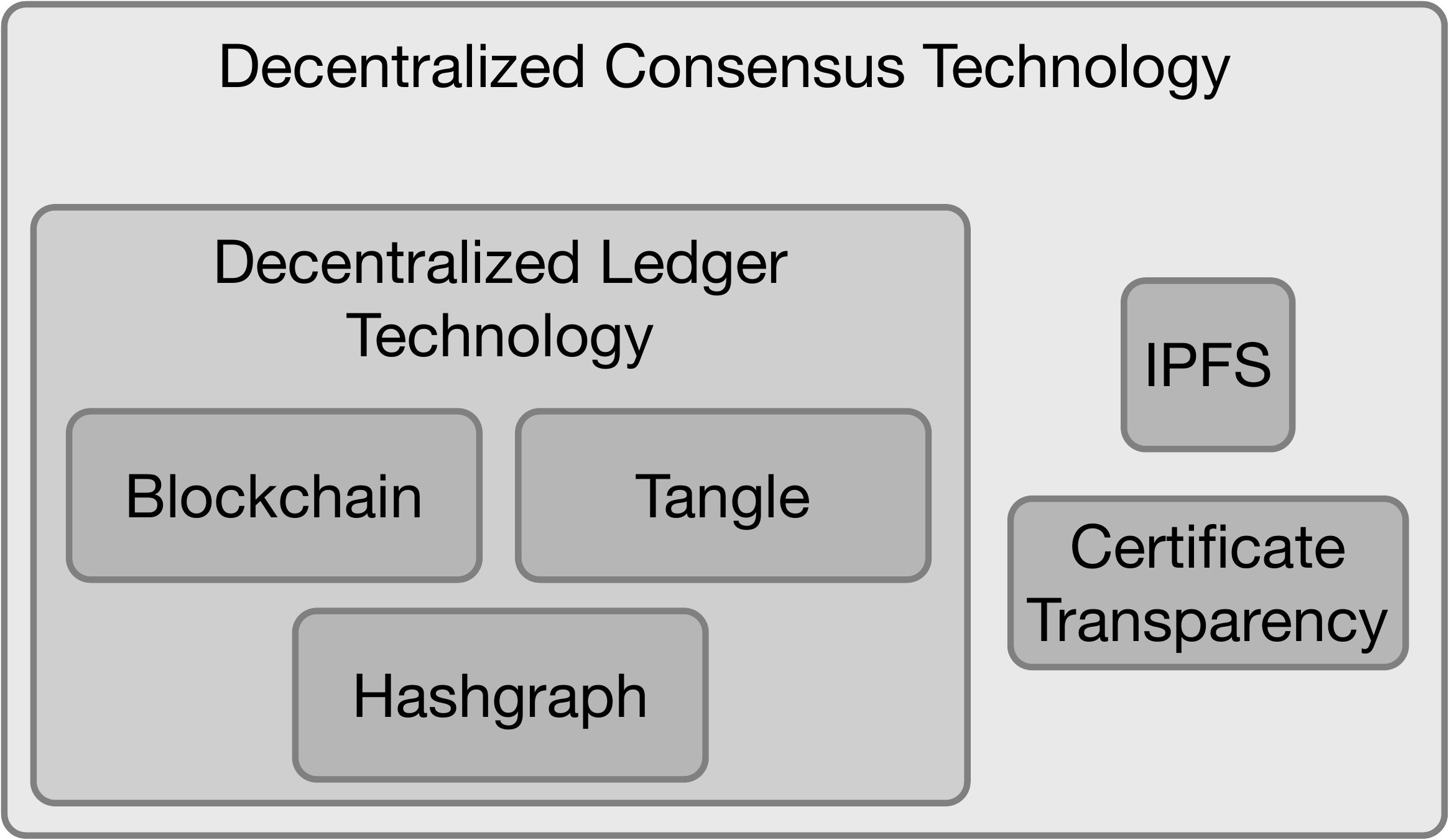}
	\caption{Decentralized Consensus Technology and its subsets\label{fig:blockchain-subsets}}
\end{figure}

Our understanding of the terminology and the relationship of the different terms is visualized in Figure~\ref{fig:blockchain-subsets}. The following Section~\ref{lab:sec:dlt} will briefly explain the differentiation and will try to give examples for the different categories. Despite the differences between the systems summarized with the term \textit{Decentralized Consensus Technology}, we still believe that there are shared properties. We will elaborate further on this in Section~\ref{lab:sec:dct:properties}.     

\subsection{Differentiation of Distributed Ledger Technology\label{lab:sec:dlt}}
We described before that a lot of projects and innovations are based on Bitcoin-like blockchain technology but sometimes modify important properties of the original concept. Nevertheless, they are very often still described with term blockchain technology. While this is probably correct in some cases, there are counterexamples for which it is plainly misleading: IOTA's Tangle for example ("Blockchain without Blocks and the Chain" \cite{schiener_primer_2017}) or Hashgraph, which clearly use different data structures and have steered away from the initial concept in important aspects. To regard this fact in literature the term "\textit{Decentralized Ledger Technology}" is sometimes used (e.g. \cite{zhang_towards_2018}). 

Nonetheless, we still deem that term to be too specific for some applications: It suggests that the members of the network in question are trying to agree on a commonly available ledger (e.g. the transactions record). On the other hand there are many projects that use a technology inspired by Bitcoin that does not necessarily have a ledger. Considering projects like IPFS \cite{benet_ipfs_2018}, respectively Filecoin \cite{protocol_labs_filecoin:_2017}, it becomes clear that there are also projects agreeing on elements other than a shared ledger. The Filecoin network, for example, agrees on a shared (virtual) file system. Even though there are projects combining these ideas with a ledger (like Storj does to create a payment and incentive system within their network \cite{wilkinson_storj_2016}), it becomes apparent that the term \textit{Distributed Ledger Technology} is too restrictive for these kinds of projects. Another example would be the Google project Certificate Transparency\footnote{\url{https://www.certificate-transparency.org/} (visited on 2018-10-19)} which aims at making SSL/TLS certificates more secure. One of the founders of the project lists several arguments why Bitcoin-like blockchain projects are not suitable for the project goals \cite{laurie_decentralised_2011}. Nevertheless, the implementation of the project has some remarkable similarities to the initial Bitcoin ideas (e.g. the Log Servers are using append-only data structures to store the monitored certificates). Even though the Log Servers do not directly agree on one global shared log, the network is able to agree on shared knowledge about which certificates have been issued officially and are valid therefore. Following this argumentation, the term \textit{Distributed Ledger Technology} is still too restrictive to cover the different projects. We propose to use the term \textit{Decentralized Consensus Technology} instead. %
The example of the Certificate Transparency project can also be used to argue, why we propose to use the term "\textit{Decentralized} Consensus Technology" instead of "\textit{Distributed} Consensus Technology": There are several projects that favor a decentralized architecture (following the definition of Baran~\cite{baran_distributed_1964}) but do not necessarily are designed in a fully distributed way (in which multiple paths exist between nearly every node; see Figure~\ref{fig:decentralized:distributed}). Further examples for systems being decentralized but not always distributed are some consortium blockchains (see Section~\ref{lab:sec:blockchain:types}).

\subsection{Properties of Decentralized Consensus Technology \label{lab:sec:dct:properties}}
In Section~\ref{lab:sec:research} we presented several projects and research approaches that are based on Nakamoto's blockchain concept and further evolved it. We showed that there is no common agreement about the terms and properties of these technologies and therefore proposed the term \textit{Decentralized Consensus Technology}. Given the presented research and literature (e.g. \cite{conte_de_leon_blockchain:_2017,zhang_towards_2018,bonneau_sok:_2015,chatterjee_overview_2017}) we propose to use the following three properties as a base line for \textit{Decentralized Consensus Technology}:   

\begin{itemize}
	\item \textbf{Decentralization:} One characteristic of most evaluated projects and research approaches was the common goal to replace a, previously central, service with a decentralized or distributed solution. This is even true for private and consortium blockchains which do not aim at including every willing participant but only some selected ones.
	\item \textbf{Trustlessness and Resilience:} Another property is the general agreement to develop a solution that helps to shift the needed trust from people or institutions to hardware- or software-based algorithms. Furthermore, all approaches we know about are based on the assumption that participants might spread, consciously or by mistake, false and fraudulent messages.  
	\item \textbf{Eventual Consensus and Growing Confidence:} The last property we observed during our studies was the common goal to agree on a shared state with all other network members. Herein, the state might be both a ledger or something abstract like the certificates that are seen as valid or the state after executing decentralized code. Furthermore, all approaches known to us, are presenting a methodology in which fraudulent modifications of already issued messages are getting harder and therefore unlikelier the older the message, respectively the block, is. 
\end{itemize}

In addition to these properties, we believe that there are further, use case specific ones. Consider a cryptocurrency that should be developed for highly privacy concerned users, in which case anonymity might also be a key property of the new system. On the other hand, if the application goal depends on smart contracts, the ability to execute code in a decentralized environment might be a core attribute of the system. Wessling et al.~\cite{wessling_how_2018} propose to use the trust property in order to determine whether and where a blockchain should be integrated into the planned application architecture. 
We propose to do this for other properties as well. 
Properties that should be considered are the following: 
\begin{itemize}
	\item \textit{Privacy} describes the degree of anonymity, respectively retraceability, of transactions in the system. Is it possible to identify the real world identity of the sender and receiver of transactions? 
	\item \textit{Participation incentive} means the way how honest behavior is incentivized and dishonest discouraged. This decision affects the used consensus algorithm, respectively is affected by the selected one.
	\item \textit{Irreversibility} and \textit{immutability} deals with whether transactions that have been accepted by the network can be revoked later on. Is it possible to revert actions unwanted by the majority of the network?
	\item \textit{Operation purpose} describes the overall goal of the system. Should it serve as a decentralized currency or should it be possible to execute arbitrary computations in a decentralized system?  
	\item \textit{Confirmation time} and \textit{Transaction costs} have to be considered when designing or selecting a suitable foundation of the application to be developed. Both affect the stability and security of the network but might greatly influence how a new application has to be used.
	\item \textit{Ability to externalize transactions and computations} enables possibilities to decrease the financial and time costs of transactions in the network. Sidechains and Off-Chaining are ways how frequent or complex transactions and calculations can be securely included in the network without negatively influencing the other participants.
	\item \textit{Scalability possibilities} are ways how the performance of the network can be increased without decreasing the overall security. Several ways have been proposed in the previous years how this can be archived but the impacts and consequences of these ways have to be considered when designing a new application using such a network.
\end{itemize}

\section{Related Work \label{lab:sec:related:work}}
At the time of writing, there are only a few papers in the context of our work. For example Bonneau et al.~\cite{bonneau_sok:_2015} and Conte de Leon et al.~\cite{conte_de_leon_blockchain:_2017} evaluate research questions and challenges in the context of cryptocurrencies. 
In contrast to this work they evaluate the strengths and weaknesses of blockchain-like systems without using this knowledge to advocate a new term describing systems that broadly differ from the initial Bitcoin concept. 
Furthermore, Conte de Leon et al.~\cite{conte_de_leon_blockchain:_2017} even use the term "distributed ledger system" throughout their paper without describing how this differs from blockchain systems, but uses both descriptions synonymously.
Both Bonneau et al.~\cite{bonneau_sok:_2015} and Conte de Leon et al.~\cite{conte_de_leon_blockchain:_2017}, of course, only consider work that was known during the time of writing of their articles. 
Since \textit{Decentralized Consensus Technology} is a fast developing field of research a lot of important aspects and improvements of the recent years are therefore not present and considered in their work but in ours.     

There are different works trying to identify properties of blockchain systems (especially Bitcoin), such as \cite{chatterjee_overview_2017}, \cite{cachin_blockchain_2017}, \cite{lin_survey_2017} or \cite{walch_path_2017}. Approaches like \cite{glaser_beyond_2015} focus a broader and less technical view on that topic and studies like \cite{zhang_towards_2018} evaluate the properties of such systems with focus on the attributes \textit{decentralisation}, \textit{consistency} and \textit{scalability} in comparison with the CAP theorem of traditional database systems. %
Furthermore, there are efforts from the National Institute of Standards and Technology (NIST) \cite{yaga_blockchain_2018} and the International Organization for Standardization (ISO)\footnote{\url{https://www.iso.org/committee/6266604.html} (visited on 2018-10-22)} to define terms in the context of \textit{Decentralized Consensus Technology}. To our knowledge, both exclusively focus on blockchain technology (like Bitcoin) and do not propose a more generic term to cope with the properties new research approaches modify, remove or add. In contrast to our contribution, their target audience has none or little knowledge of blockchain technology (cf. \cite[p. iii]{yaga_blockchain_2018}).

\section{Conclusion \label{lab:sec:conclusion}}
Starting with the Bitcoin concept a lot of research has been done in the context of blockchain technology and applications. Unfortunately, there is no commonly shared knowledge about terms in that field of research. Various research papers start with a (new) introduction to that topic and partly new terms. The core concept as implemented by Bitcoin is often called blockchain technology because (financial) transactions are bundled in blocks which refer to their ancestors and therefore form a chain. Due to their cryptographic implementation and game theoretical concepts, the confidence that a certain block cannot be modified, grows over time. The initial concept by Nakamoto has some problems, like scalability issues, attack vectors or confirmation time and costs for new transactions. In the last years a lot of research, both by academical and industrial researchers, has been done to address these problems. Some of them modify the core principles of the original concept significantly or focus on other properties than the initial proposal. It has to be questioned whether these new concepts still adequately could be described as blockchain technology. Consider technologies explicitly refusing to be seen as blockchain technology but following similar ideas. We proposed to use the term \textit{Decentralized Consensus Technology} as a general category instead. \textit{Decentralized Consensus Technology} consists of decentralized ledger and non-ledger technologies. Blockchain technology in turn is only one variant of \textit{Decentralized Ledger Technology}.  

Furthermore, we identified three main characteristics of \textit{Decentralized Consensus Technologies}: \textit{decentralization}, \textit{trustlessness} and its \textit{ability to eventually reach consensus}. Depending on the use case of the specific implementation other properties also have to be considered. We gave an overview of the current research and showed which properties of the initial concept are modified or removed and which new attributes are added by the various scientific contributions. We used this insights to further derive use case specific properties like \textit{privacy}, \textit{participation incentive}, \textit{irreversibility} and \textit{immutability}, \textit{operation purpose}, \textit{confirmation time}, \textit{transaction costs}, \textit{ability to externalize transactions and computations} and \textit{scalability possibilities}. We propose to take these terminology and properties into consideration when evaluating, describing and modelling the creation or usage of \textit{Decentralized Consensus Technology} in upcoming projects and research.

\bibliographystyle{splncs04}
\bibliography{bibliography}

\begin{thebibliography}{100}
\providecommand{\url}[1]{\texttt{#1}}
\providecommand{\urlprefix}{URL }
\providecommand{\doi}[1]{https://doi.org/#1}

\bibitem{andreas_m._antonopoulos_introduction_2016}
{Andreas M. Antonopoulos}: Introduction to bitcoin (2016),
  \url{https://www.youtube.com/watch?v=l1si5ZWLgy0}

\bibitem{androulaki_hiding_2014}
Androulaki, E., Karame, G.O.: Hiding transaction amounts and balances in
  bitcoin. In: Holz, T., Ioannidis, S. (eds.) Trust and Trustworthy Computing.
  pp. 161--178. Lecture Notes in Computer Science, Springer International
  Publishing (2014)

\bibitem{antonopoulos_mastering_2015}
Antonopoulos, A.M.: Mastering bitcoin. O'Reilly, 1. edition edn. (2015)

\bibitem{antonopoulos_mastering_2017}
Antonopoulos, A.M.: Mastering Bitcoin: Programming the Open Blockchain.
  "O'Reilly Media, Inc." (2017)

\bibitem{antonopoulos_mastering_2018}
Antonopoulos, A.M., Wood, G.: Mastering Ethereum: Building Smart Contracts and
  Dapps. O'Reilly Media, Incorporated (2018),
  \url{https://github.com/ethereumbook/ethereumbook}, google-Books-{ID}:
  {SedSMQAACAAJ}

\bibitem{armknecht_ripple:_2015}
Armknecht, F., Karame, G.O., Mandal, A., Youssef, F., Zenner, E.: Ripple:
  Overview and outlook. In: Conti, M., Schunter, M., Askoxylakis, I. (eds.)
  Trust and Trustworthy Computing. pp. 163--180. Lecture Notes in Computer
  Science, Springer International Publishing (2015)

\bibitem{aru_can_2018}
Aru, I.: Can blockchain technology survive without cryptocurrencies? (2018),
  \url{https://www.ccn.com/can-blockchain-technology-survive-without-cryptocurrencies/}

\bibitem{ateniese_redactable_2017}
Ateniese, G., Magri, B., Venturi, D., Andrade, E.: Redactable blockchain - or -
  rewriting history in bitcoin and friends. In: 2017 {IEEE} European Symposium
  on Security and Privacy ({EuroS} P). pp. 111--126 (2017).
  \doi{10.1109/EuroSP.2017.37}

\bibitem{atzei_survey_2017}
Atzei, N., Bartoletti, M., Cimoli, T.: A survey of attacks on ethereum smart
  contracts ({SoK}). In: Maffei, M., Ryan, M. (eds.) Principles of Security and
  Trust. pp. 164--186. Lecture Notes in Computer Science, Springer Berlin
  Heidelberg (2017)

\bibitem{azaria_medrec:_2016}
Azaria, A., Ekblaw, A., Vieira, T., Lippman, A.: {MedRec}: Using blockchain for
  medical data access and permission management. In: 2016 2nd International
  Conference on Open and Big Data ({OBD}). pp. 25--30 (2016).
  \doi{10.1109/OBD.2016.11}

\bibitem{back_hashcash}
Back, A.: Hashcash - a denial of service counter-measure (2002),
  \url{http://www.hashcash.org/hashcash.pdf}

\bibitem{back_enabling_2014}
Back, A., Corallo, M., Dashjr, L., Friedenbach, M., Maxwell, G., Miller, A.,
  Poelstra, A., Tim\'{o}n, J., Wuille, P.: Enabling blockchain innovations with
  pegged sidechains (2014), \url{http://kevinriggen.com/files/sidechains.pdf}

\bibitem{badertscher_but_2018}
Badertscher, C., Garay, J., Maurer, U., Tschudi, D., Zikas, V.: But why does it
  work? a rational protocol design treatment of bitcoin. In: Advances in
  Cryptology – {EUROCRYPT} 2018. pp. 34--65. Lecture Notes in Computer
  Science, Springer, Cham (2018). \doi{10.1007/978-3-319-78375-8\_2},
  \url{https://link.springer.com/chapter/10.1007/978-3-319-78375-8\_2}

\bibitem{baird_swirlds_2016}
Baird, L.: The swirlds hashgraph consensus algorithm: Fair, fast, byzantine
  fault tolerance (2016),
  \url{https://www.swirlds.com/downloads/SWIRLDS-TR-2016-01.pdf}

\bibitem{baird_methods_2017}
Baird, L.: Methods and apparatus for a distributed database within a network
  (2017), \url{https://patents.google.com/patent/US9646029/en}

\bibitem{bamert_have_2013}
Bamert, T., Decker, C., Elsen, L., Wattenhofer, R., Welten, S.: Have a snack,
  pay with bitcoins. In: 13th {IEEE} International Conference on Peer-to-Peer
  Computing. pp.~1--5 (2013). \doi{10.1109/P2P.2013.6688717}

\bibitem{baran_distributed_1964}
Baran, P.: On distributed communications networks. {IEEE} Transactions on
  Communications Systems  \textbf{12}(1), ~1--9 (1964).
  \doi{10.1109/TCOM.1964.1088883}

\bibitem{bartoletti_analysis_2017}
Bartoletti, M., Pompianu, L.: An analysis of bitcoin {OP}\_return metadata. In:
  Financial Cryptography and Data Security. pp. 218--230. Lecture Notes in
  Computer Science, Springer International Publishing (2017)

\bibitem{bartolucci_sharvot:_2018}
Bartolucci, S., Bernat, P., Joseph, D.: {SHARVOT}: Secret {SHARe}-based
  {VOTing} on the blockchain. In: Proceedings of the 1st International Workshop
  on Emerging Trends in Software Engineering for Blockchain. pp. 30--34.
  {WETSEB} '18, {ACM} (2018). \doi{10.1145/3194113.3194118},
  \url{http://doi.acm.org/10.1145/3194113.3194118}

\bibitem{beck_blockchain_2016}
Beck, R., Czepluch, J.S., Lollike, N., Malone, S.: Blockchain - the gateway to
  trust-free crypthographic transactions. In: Twenty-Fourth European Conference
  on Information Systems ({ECIS}). p.~15 (2016)

\bibitem{beikverdi_trend_2015}
Beikverdi, A., {JooSeok Song}: Trend of centralization in bitcoin's distributed
  network. In: 2015 {IEEE}/{ACIS} 16th International Conference on Software
  Engineering, Artificial Intelligence, Networking and Parallel/Distributed
  Computing ({SNPD}). pp.~1--6. {IEEE} (2015). \doi{10.1109/SNPD.2015.7176229},
  \url{http://ieeexplore.ieee.org/document/7176229/}

\bibitem{belizaire_soluna_2018}
Belizaire, J.: Soluna powering the blockchain (2018),
  \url{https://soluna.io/wp-content/uploads/2018/09/Soluna_white_paper_v1.2_20180905.pdf}

\bibitem{ben-or_completeness_1988}
Ben-Or, M., Goldwasser, S., Wigderson, A.: Completeness theorems for
  non-cryptographic fault-tolerant distributed computation. In: Proceedings of
  the Twentieth Annual {ACM} Symposium on Theory of Computing. pp. 1--10.
  {STOC} '88, {ACM} (1988). \doi{10.1145/62212.62213},
  \url{http://doi.acm.org/10.1145/62212.62213}

\bibitem{benet_ipfs_2018}
Benet, J.: {IPFS} - content addressed, versioned, p2p file system (2018),
  \url{https://ipfs.io/ipfs/QmR7GSQM93Cx5eAg6a6yRzNde1FQv7uL6X1o4k7zrJa3LX/ipfs.draft3.pdf}

\bibitem{bentov_cryptocurrencies_2016}
Bentov, I., Gabizon, A., Mizrahi, A.: Cryptocurrencies without proof of work.
  In: Financial Cryptography and Data Security. pp. 142--157. Lecture Notes in
  Computer Science, Springer, Berlin, Heidelberg (2016).
  \doi{10.1007/978-3-662-53357-4\_10},
  \url{https://link.springer.com/chapter/10.1007/978-3-662-53357-4\_10}

\bibitem{bentov_proof_2014}
Bentov, I., Lee, C., Mizrahi, A., Rosenfeld, M.: Proof of activity: Extending
  bitcoin's proof of work via proof of stake [extended abstract]. {ACM}
  {SIGMETRICS} Performance Evaluation Review  \textbf{42}(3),  34--37 (2014).
  \doi{10.1145/2695533.2695545},
  \url{http://doi.acm.org/10.1145/2695533.2695545}

\bibitem{bocek_smart_2018}
Bocek, T., Stiller, B.: Smart contracts - blockchains in the wings. In:
  Linnhoff-Popien, C., Schneider, R., Zaddach, M. (eds.) Digital Marketplaces
  Unleashed, pp. 169--184. Springer Berlin Heidelberg (2018).
  \doi{10.1007/978-3-662-49275-8\_19},
  \url{https://doi.org/10.1007/978-3-662-49275-8\_19}

\bibitem{bodrova_what_2017}
Bodrova, A.: What are decentralized applications (dapps) (2017),
  \url{https://medium.com/ethereum-dapp-builder/what-are-decentralized-applications-dapps-ed7459a27786}

\bibitem{boehm_bitcoin:_2014}
Boehm, F., Pesch, P.: Bitcoin: A first legal analysis. In: Financial
  Cryptography and Data Security. pp. 43--54. Lecture Notes in Computer
  Science, Springer, Berlin, Heidelberg (2014).
  \doi{10.1007/978-3-662-44774-1\_4},
  \url{https://link.springer.com/chapter/10.1007/978-3-662-44774-1\_4}

\bibitem{bohr_who_2014}
Bohr, J., Bashir, M.: Who uses bitcoin? an exploration of the bitcoin
  community. In: 2014 Twelfth Annual International Conference on Privacy,
  Security and Trust. pp. 94--101 (2014). \doi{10.1109/PST.2014.6890928}

\bibitem{bonneau_sok:_2015}
Bonneau, J., Miller, A., Clark, J., Narayanan, A., Kroll, J.A., Felten, E.W.:
  {SoK}: Research perspectives and challenges for bitcoin and cryptocurrencies.
  In: 2015 {IEEE} Symposium on Security and Privacy. pp. 104--121 (2015).
  \doi{10.1109/SP.2015.14}

\bibitem{bonneau_mixcoin:_2014}
Bonneau, J., Narayanan, A., Miller, A., Clark, J., Kroll, J.A., Felten, E.W.:
  Mixcoin: Anonymity for bitcoin with accountable mixes. In: Christin, N.,
  Safavi-Naini, R. (eds.) Financial Cryptography and Data Security. pp.
  486--504. Lecture Notes in Computer Science, Springer Berlin Heidelberg
  (2014)

\bibitem{bradbury_blocks_2015}
Bradbury, D.: In blocks [security bitcoin]. Engineering Technology
  \textbf{10}(2),  68--71 (2015). \doi{10.1049/et.2015.0208}

\bibitem{buterin_ethereum_2014}
Buterin, V.: Ethereum {White} {Paper} - {A} {Next}-{Generation} {Smart}
  {Contract} and {Decentralized} {Application} {Platform} (2014),
  \url{https://github.com/ethereum/wiki}

\bibitem{buterin_meaning_2017}
Buterin, V.: The meaning of decentralization (2017),
  \url{https://medium.com/@VitalikButerin/the-meaning-of-decentralization-a0c92b76a274}

\bibitem{buterin_prehistory_2017}
Buterin, V.: A prehistory of the ethereum protocol (2017),
  \url{https://vitalik.ca/general/2017/09/14/prehistory.html}

\bibitem{buterin_casper_2017}
Buterin, V., Griffith, V.: Casper the friendly finality gadget.
  {arXiv}:1710.09437 [cs]  (2017), \url{http://arxiv.org/abs/1710.09437}

\bibitem{cachin_blockchain_2017}
Cachin, C.: Blockchain, cryptography, and consensus (2017),
  \url{https://www.itu.int/en/ITU-T/Workshops-and-Seminars/201703/Documents/Christian%20Cachin%20blockchain-itu.pdf},
  {ITU} Workshop on "Security Aspects of Blockchain"

\bibitem{carlsten_instability_2016}
Carlsten, M., Kalodner, H., Weinberg, S.M., Narayanan, A.: On the instability
  of bitcoin without the block reward. In: Proceedings of the 2016 {ACM}
  {SIGSAC} Conference on Computer and Communications Security. pp. 154--167.
  {CCS} '16, {ACM} (2016). \doi{10.1145/2976749.2978408},
  \url{http://doi.acm.org/10.1145/2976749.2978408}

\bibitem{castro_practical_1999}
Castro, M., Liskov, B.: Practical byzantine fault tolerance. In: Proceedings of
  the Third Symposium on Operating Systems Design and Implementation. vol.~99,
  pp. 173--186 (1999)

\bibitem{chatterjee_overview_2017}
Chatterjee, R.: An overview of the emerging technology: Blockchain. In: 2017
  3rd International Conference on Computational Intelligence and Networks
  ({CINE}). pp. 126--127 (2017). \doi{10.1109/CINE.2017.33}

\bibitem{chen_security_2017}
Chen, L., Xu, L., Shah, N., Gao, Z., Lu, Y., Shi, W.: On security analysis of
  proof-of-elapsed-time ({PoET}). In: Stabilization, Safety, and Security of
  Distributed Systems. pp. 282--297. Lecture Notes in Computer Science,
  Springer, Cham (2017). \doi{10.1007/978-3-319-69084-1\_19},
  \url{https://link.springer.com/chapter/10.1007/978-3-319-69084-1\_19}

\bibitem{chen_efficient_1992}
Chen, M., Wu, K., Yu, P.S.: Efficient decentralized consensus protocols in a
  distributed computing system. In: [1992] Proceedings of the 12th
  International Conference on Distributed Computing Systems. pp. 426--433
  (1992). \doi{10.1109/ICDCS.1992.235012}

\bibitem{chohan_cryptocurrency_2017}
Chohan, U.W.: The cryptocurrency tumblers: Risks, legality and oversight.
  {SSRN} Scholarly Paper {ID} 3080361, Social Science Research Network (2017),
  \url{https://papers.ssrn.com/abstract=3080361}

\bibitem{noauthor_protocol_2018}
counterparty.io: Protocol specification {\textbar} counterparty (2018),
  \url{https://counterparty.io/docs/protocol_specification/}

\bibitem{dai_b-money_1998}
Dai, W.: b-money (1998), \url{http://www.weidai.com/bmoney.txt}

\bibitem{decker_information_2013}
Decker, C., Wattenhofer, R.: Information propagation in the bitcoin network.
  In: {IEEE} P2P 2013 Proceedings. pp. 1--10 (2013).
  \doi{10.1109/P2P.2013.6688704}

\bibitem{duffy_everything_2018}
Duffy, J.M.: Everything you need to know about loom network, all in one place
  (updated regularly) (2018),
  \url{https://medium.com/loom-network/everything-you-need-to-know-about-loom-network-all-in-one-place-updated-regularly-64742bd839fe}

\bibitem{dwork_pricing_1993}
Dwork, C., Naor, M.: Pricing via processing or combatting junk mail. In:
  Brickell, E.F. (ed.) Advances in Cryptology — {CRYPTO}’ 92. pp. 139--147.
  Lecture Notes in Computer Science, Springer Berlin Heidelberg (1993)

\bibitem{eberhardt_or_2017}
Eberhardt, J., Tai, S.: On or off the blockchain? insights on off-chaining
  computation and data. In: Service-Oriented and Cloud Computing. pp. 3--15.
  Lecture Notes in Computer Science, Springer, Cham (2007).
  \doi{10.1007/978-3-319-67262-5\_1},
  \url{https://link.springer.com/chapter/10.1007/978-3-319-67262-5\_1}

\bibitem{ehmke_proof--property:_2018}
Ehmke, C., Wessling, F., Friedrich, C.M.: Proof-of-property: a lightweight and
  scalable blockchain protocol. In: Proceedings of the 1st International
  Workshop on Emerging Trends in Software Engineering for Blockchain. pp.
  48--51. {ACM} Press (2018). \doi{10.1145/3194113.3194122},
  \url{http://dl.acm.org/citation.cfm?doid=3194113.3194122}

\bibitem{ekblaw_bitcoin_2016}
Ekblaw, A., Barabas, C., Harvey-Buschel, J., Lippman, A.: Bitcoin and the myth
  of decentralization: Socio-technical proposals for restoring network
  integrity. In: 2016 {IEEE} 1st International Workshops on Foundations and
  Applications of Self* Systems ({FAS}*W). pp. 18--23 (2016).
  \doi{10.1109/FAS-W.2016.18}

\bibitem{eyal_bitcoin-ng:_2016}
Eyal, I., Gencer, A.E., Sirer, E.G., Van~Renesse, R.: Bitcoin-{NG}: A scalable
  blockchain protocol. In: Proceedings of the 13th Usenix Conference on
  Networked Systems Design and Implementation. pp. 45--59. {NSDI}'16, {USENIX}
  Association (2016), \url{http://dl.acm.org/citation.cfm?id=2930611.2930615}

\bibitem{eyal_majority_2018}
Eyal, I., Sirer, E.G.: Majority is not enough: bitcoin mining is vulnerable.
  Communications of the {ACM}  \textbf{61}(7),  95--102 (2018).
  \doi{10.1145/3212998},
  \url{http://dl.acm.org/citation.cfm?doid=3234519.3212998}

\bibitem{finestone_game_2017}
Finestone, M.: Game theory and blockchain --- coinsquare news (2017),
  \url{https://news.coinsquare.com/digital-currency/game-theory-and-blockchain/}

\bibitem{fischer_impossibility_1985}
Fischer, M.J., Lynch, N.A., Paterson, M.S.: Impossibility of distributed
  consensus with one faulty process. Journal of the {ACM} ({JACM})
  \textbf{32}(2),  374--382 (1985). \doi{10.1145/3149.214121},
  \url{http://doi.acm.org/10.1145/3149.214121}

\bibitem{foley_sex_2018}
Foley, S., Karlsen, J.R., Putni{\c{n}}{\v{s}}, T.J.: Sex, drugs, and bitcoin:
  How much illegal activity is financed through cryptocurrencies? {SSRN}
  Electronic Journal  (2018). \doi{10.2139/ssrn.3102645},
  \url{https://www.ssrn.com/abstract=3102645}

\bibitem{franco_understanding_2015}
Franco, P.: Understanding bitcoin: cryptography, engineering and economics.
  John Wiley \& Sons (2015),
  \url{http://www.books24x7.com/marc.asp?bookid=80664}, {OCLC}: 894170560

\bibitem{gao_two_2016}
Gao, X., Clark, G.D., Lindqvist, J.: Of two minds, multiple addresses, and one
  ledger: Characterizing opinions, knowledge, and perceptions of bitcoin across
  users and non-users. In: Proceedings of the 2016 {CHI} Conference on Human
  Factors in Computing Systems. pp. 1656--1668. {CHI} '16, {ACM} (2016).
  \doi{10.1145/2858036.2858049},
  \url{http://doi.acm.org/10.1145/2858036.2858049}

\bibitem{gencer_decentralization_2018}
Gencer, A.E., Basu, S., Eyal, I., Sirer, E.G.: Decentralization in bitcoin and
  ethereum networks. In: Financial Cryptography and Data Security 2018. p.~18
  (2018)

\bibitem{gervais_is_2014}
Gervais, A., Karame, G.O., Capkun, V., Capkun, S.: Is bitcoin a decentralized
  currency? {IEEE} Security \& Privacy  \textbf{12}(3),  54--60 (2014).
  \doi{10.1109/MSP.2014.49},
  \url{http://ieeexplore.ieee.org/lpdocs/epic03/wrapper.htm?arnumber=6824541}

\bibitem{gervais_tampering_2015}
Gervais, A., Ritzdorf, H., Karame, G.O., Capkun, S.: Tampering with the
  delivery of blocks and transactions in bitcoin. In: Proceedings of the 22Nd
  {ACM} {SIGSAC} Conference on Computer and Communications Security. pp.
  692--705. {CCS} '15, {ACM} (2015). \doi{10.1145/2810103.2813655},
  \url{http://doi.acm.org/10.1145/2810103.2813655}

\bibitem{gipp_decentralized_2015}
Gipp, B., Meuschke, N., Gernandt, A.: Decentralized trusted timestamping using
  the crypto currency bitcoin (2015),
  \url{https://www.gipp.com/wp-content/papercite-data/pdf/gipp15a.pdf}

\bibitem{glaser_beyond_2015}
Glaser, F., Bezzenberger, L.: Beyond cryptocurrencies - a taxonomy of
  decentralized consensus systems. In: 23rd European Conference on Information
  Systems ({ECIS}). p.~19 (2015)

\bibitem{goranovic_blockchain_2017}
Goranovi\'{c}, A., Meisel, M., Fotiadis, L., Wilker, S., Treytl, A., Sauter,
  T.: Blockchain applications in microgrids an overview of current projects and
  concepts. In: {IECON} 2017 - 43rd Annual Conference of the {IEEE} Industrial
  Electronics Society. pp. 6153--6158 (2017). \doi{10.1109/IECON.2017.8217069}

\bibitem{grier_all_2014}
Grier, D.A.: All that glitters is not gold. Computer  \textbf{47}(4),  116--116
  (2014). \doi{10.1109/MC.2014.79}

\bibitem{gries_using_2018}
Gries, S., Meyer, O., Wessling, F., Hesenius, M., Gruhn, V.: Using blockchain
  technology to ensure trustful information flow monitoring in {CPS}. In: 2018
  {IEEE} International Conference on Software Architecture Companion
  ({ICSA}-C). pp. 35--38 (2018). \doi{10.1109/ICSA-C.2018.00014}

\bibitem{heilman_eclipse_2015}
Heilman, E., Kendler, A., Zohar, A., Goldberg, S.: Eclipse attacks on bitcoin's
  peer-to-peer network. In: Proceedings of the 24th {USENIX} Conference on
  Security Symposium. pp. 129--144. {SEC}'15, {USENIX} Association (2015),
  \url{http://dl.acm.org/citation.cfm?id=2831143.2831152}

\bibitem{hern_2014}
Hern, A.: Bitcoin currency could have been destroyed by '51
  2014),
  \url{https://www.theguardian.com/technology/2014/jun/16/bitcoin-currency-destroyed-51-attack-ghash-io}

\bibitem{hornyak_2014}
Hornyak, T.: One group controls 51 percent of bitcoin mining, threatening
  security sanctity (Jun 2014),
  \url{https://www.pcworld.com/article/2364000/bitcoin-price-dips-as-backers-fear-mining-monopoly.html}

\bibitem{hrones_segwit_2017}
Hrones, M.: Segwit activated: How it works \& what's next for bitcoin (2017),
  \url{https://bitcoinist.com/segwit-activated-next-bitcoin/}

\bibitem{iansiti_truth_2017}
Iansiti, M., Lakhani, K.R.: The truth about blockchain. Harvard Business Review
   \textbf{95}(1),  118--127 (2017),
  \url{https://hbr.org/2017/01/the-truth-about-blockchain}

\bibitem{johnston_decentralizedapplications}
Johnston, D., Yilmaz, S.O., Kandah, J., Bentenitis, N., Hashemi, F., Gross, R.,
  Wilkinson, S., Mason, S.: {DecentralizedApplications}: Decentralized
  applications white paper and spec (2013),
  \url{https://github.com/DavidJohnstonCEO/DecentralizedApplications}

\bibitem{karame_two_2012}
Karame, G.O.: Two bitcoins at the price of one? double-spending attacks on fast
  payments in bitcoin. In: In Proc. of Conference on Computer and Communication
  Security (2012)

\bibitem{kiffer_stick_2017}
Kiffer, L., Levin, D., Mislove, A.: Stick a fork in it: Analyzing the ethereum
  network partition. In: Proceedings of the 16th {ACM} Workshop on Hot Topics
  in Networks. pp. 94--100. {HotNets}-{XVI}, {ACM} (2017).
  \doi{10.1145/3152434.3152449},
  \url{http://doi.acm.org/10.1145/3152434.3152449}

\bibitem{king_primecoin:_2013}
King, S.: Primecoin: Cryptocurrency with prime number proof-of-work (2013),
  \url{http://www.cryptocoin.kr/attachment/cfile7.uf@99A6C1365AF7044D238ED8.pdf}

\bibitem{kokoris-kogias_omniledger_2018}
Kokoris-Kogias, E., Jovanovic, P., Gasser, L., Gailly, N., Syta, E., Ford, B.:
  {OmniLedger}: A secure, scale-out, decentralized ledger via sharding. {IEEE}
  Symposium on Security and Privacy ({SP}) pp. 19--34 (2018).
  \doi{10.1109/SP.2018.000-5}

\bibitem{kravchenko_does_2016}
Kravchenko, P.: Does a blockchain really need a native coin? (2016),
  \url{https://medium.com/@pavelkravchenko/does-a-blockchain-really-need-a-native-coin-f6a5ff2a13a3}

\bibitem{kshetri_can_2017}
Kshetri, N.: Can blockchain strengthen the internet of things? {IT}
  Professional  \textbf{19}(4),  68--72 (2017). \doi{10.1109/MITP.2017.3051335}

\bibitem{kumar_traceability_2017}
Kumar, A., Fischer, C., Tople, S., Saxena, P.: A traceability analysis of
  monero's blockchain. In: Foley, S.N., Gollmann, D., Snekkenes, E. (eds.)
  Computer Security - {ESORICS} 2017. pp. 153--173. Lecture Notes in Computer
  Science, Springer International Publishing (2017)

\bibitem{Kunnapas2016}
K{\"u}nnapas, K.: From Bitcoin to Smart Contracts: Legal Revolution or
  Evolution from the Perspective of de lege ferenda?, pp. 111--131. Springer
  International Publishing, Cham (2016). \doi{10.1007/978-3-319-26896-5\_6},
  \url{https://doi.org/10.1007/978-3-319-26896-5\_6}

\bibitem{protocol_labs_filecoin:_2017}
Labs, P.: Filecoin: A decentralized storage network (2017),
  \url{https://filecoin.io/filecoin.pdf}

\bibitem{laurie_decentralised_2011}
Laurie, B.: Decentralised currencies are probably impossible (2011),
  \url{https://www.links.org/files/decentralised-currencies.pdf}

\bibitem{lee_2017}
Lee, T.B.: A brief history of bitcoin hacks and frauds (Dec 2017),
  \url{https://arstechnica.com/tech-policy/2017/12/a-brief-history-of-bitcoin-hacks-and-frauds/}

\bibitem{conte_de_leon_blockchain:_2017}
Conte~de Leon, D., Stalick, A.Q., Jillepalli, A.A., Haney, M.A., Sheldon, F.T.:
  Blockchain: properties and misconceptions. Asia Pacific Journal of Innovation
  and Entrepreneurship  \textbf{11}(3),  286--300 (2017).
  \doi{10.1108/APJIE-12-2017-034},
  \url{http://www.emeraldinsight.com/doi/10.1108/APJIE-12-2017-034}

\bibitem{lerner_rsk_2015}
Lerner, S.D.: {RSK} - white paper overview (2015),
  \url{https://docs.rsk.co/RSK_White_Paper-Overview.pdf}

\bibitem{lewenberg_bitcoin_2015}
Lewenberg, Y., Bachrach, Y., Sompolinsky, Y., Zohar, A., Rosenschein, J.S.:
  Bitcoin mining pools: A cooperative game theoretic analysis. In: Proceedings
  of the 2015 International Conference on Autonomous Agents and Multiagent
  Systems. pp. 919--927. {AAMAS} '15, International Foundation for Autonomous
  Agents and Multiagent Systems (2015),
  \url{http://dl.acm.org/citation.cfm?id=2772879.2773270}

\bibitem{li_practical_2017}
Li, K., Yang, R., Au, M.H., Xu, Q.: Practical range proof for cryptocurrency
  monero with provable security. In: Qing, S., Mitchell, C., Chen, L., Liu, D.
  (eds.) Information and Communications Security ({ICICS} 2017). pp. 255--262.
  Lecture Notes in Computer Science, Springer International Publishing (2017)

\bibitem{noauthor_oraclize_2018}
Limited, O.: Oraclize documentation (2018),
  \url{https://docs.oraclize.it/#home}

\bibitem{lin_survey_2017}
Lin, I.C., Liao, T.C.: A survey of blockchain security issues and challenges.
  International Journal of Network Security  \textbf{19}(5),  653--659 (2017).
  \doi{10.6633/IJNS.201709.19(5).01}

\bibitem{lo_evaluating_2017}
Lo, S.K., Xu, X., Chiam, Y.K., Lu, Q.: Evaluating suitability of applying
  blockchain. In: 2017 22nd International Conference on Engineering of Complex
  Computer Systems ({ICECCS}). pp. 158--161 (2017).
  \doi{10.1109/ICECCS.2017.26}

\bibitem{lustig_algorithmic_2015}
Lustig, C., Nardi, B.: Algorithmic authority: The case of bitcoin. In: 2015
  48th Hawaii International Conference on System Sciences. pp. 743--752. {IEEE}
  (2015). \doi{10.1109/HICSS.2015.95},
  \url{http://ieeexplore.ieee.org/document/7069744/}

\bibitem{matzutt_quantitative_2018}
Matzutt, R., Hiller, J., Henze, M., Ziegeldorf, J.H., Hohlfeld, O., Wehrle, K.:
  A quantitative analysis of the impact of arbitrary blockchain content on
  bitcoin. In: Proceedings of the 22nd International Conference on Financial
  Cryptography and Data Security 2018. p.~18. Springer (2018),
  \url{https://fc18.ifca.ai/preproceedings/6.pdf}

\bibitem{merkle_protocols_1980}
Merkle, R.C.: Protocols for public key cryptosystems. In: 1980 {IEEE} Symposium
  on Security and Privacy. pp. 122--122 (1980). \doi{10.1109/SP.1980.10006}

\bibitem{miers_zerocoin:_2013}
Miers, I., Garman, C., Green, M., Rubin, A.D.: Zerocoin: Anonymous distributed
  e-cash from bitcoin. In: 2013 {IEEE} Symposium on Security and Privacy. pp.
  397--411 (2013). \doi{10.1109/SP.2013.34}

\bibitem{moser_trends_2015}
M\"oser, M., B\"ohme, R.: Trends, tips, tolls: A longitudinal study of bitcoin
  transaction fees. In: Financial Cryptography and Data Security. pp. 19--33.
  Lecture Notes in Computer Science, Springer Berlin Heidelberg (2015)

\bibitem{nakamoto_bitcoin:_2008}
Nakamoto, S.: Bitcoin: A peer-to-peer electronic cash system (2008),
  \url{https://bitcoin.org/bitcoin.pdf}, visited on 2018-10-22

\bibitem{olnes_blockchain_2017}
{\O}lnes, S., Ubacht, J., Janssen, M.: Blockchain in government: Benefits and
  implications of distributed ledger technology for information sharing.
  Government Information Quarterly  \textbf{34}(3),  355--364 (2017).
  \doi{10.1016/j.giq.2017.09.007},
  \url{http://www.sciencedirect.com/science/article/pii/S0740624X17303155}

\bibitem{olson_sawtooth:_2018}
Olson, K., Bowman, M., Mitchell, J., Middleton, D., Montgomery, C.: Sawtooth:
  An introduction (2018),
  \url{https://www.hyperledger.org/wp-content/uploads/2018/01/Hyperledger_Sawtooth_WhitePaper.pdf}

\bibitem{panetta:2017}
Panetta, K.: Top trends in the gartner hype cycle for emerging technologies,
  2017 (2017),
  \url{https://www.gartner.com/smarterwithgartner/top-trends-in-the-gartner-hype-cycle-for-emerging-technologies-2017/}

\bibitem{patel_blockchain_2017}
Patel, D., Bothra, J., Patel, V.: Blockchain exhumed. In: 2017 {ISEA} Asia
  Security and Privacy ({ISEASP}). pp. 1--12 (2017).
  \doi{10.1109/ISEASP.2017.7976993}

\bibitem{patterson_alternatives_2016}
Patterson, R.: Alternatives for proof of work, part 2: Proof of activity, proof
  of burn, proof of capacity, and byzantine generals — bytecoin blog (2016),
  \url{http://web.archive.org/web/20160304055454/https://bytecoin.org/blog/proof-of-activity-proof-of-burn-proof-of-capacity/}

\bibitem{peck_cryptographers_2018}
Peck, M.: Cryptographers urge people to abandon {IOTA} after leaked emails,
  \url{https://spectrum.ieee.org/tech-talk/computing/networks/cryptographers-urge-users-and-researchers-to-abandon-iota-after-leaked-emails}

\bibitem{peck_bitcoin:_2012}
Peck, M.E.: Bitcoin: The cryptoanarchists’ answer to cash (2012),
  \url{https://spectrum.ieee.org/computing/software/bitcoin-the-cryptoanarchists-answer-to-cash}

\bibitem{peck_blockchain_2017}
Peck, M.E.: Blockchain world - do you need a blockchain? this chart will tell
  you if the technology can solve your problem. {IEEE} Spectrum
  \textbf{54}(10),  38--60 (2017). \doi{10.1109/MSPEC.2017.8048838}

\bibitem{de_pedro_crespo_stampery_2017}
de~Pedro~Crespo, A.S., Cuende~Garc\'{i}a, L.I.: Stampery blockchain
  timestamping architecture ({BTA}) - version 6 (2017),
  \url{https://s3.amazonaws.com/stampery-cdn/docs/Stampery-BTA-v6-whitepaper.pdf}

\bibitem{perez_does_2017}
Perez, S.: Does a blockchain need a token?,
  \url{https://medium.com/swlh/does-a-blockchain-need-a-token-66c894d566fb}

\bibitem{poelstra_distributed_2014}
Poelstra, A.: Distributed consensus from proof of stake is impossible (2014),
  \url{https://download.wpsoftware.net/bitcoin/old-pos.pdf}

\bibitem{poon_plasma:_2017}
Poon, J., Buterin, V.: Plasma: Scalable autonomous smart contracts (2017),
  \url{https://plasma.io/plasma.pdf}

\bibitem{poon_bitcoin_2016}
Poon, J., Dryja, T.: The bitcoin lightning network: Scalable off-chain instant
  payments (2016), \url{https://lightning.network/lightning-network-paper.pdf}

\bibitem{pop_blockchain_2018}
Pop, C., Cioara, T., Antal, M., Anghel, I., Salomie, I., Bertoncini, M.:
  Blockchain based decentralized management of demand response programs in
  smart energy grids. Sensors  \textbf{18}(1) (2018). \doi{10.3390/s18010162}

\bibitem{popov_tangle_2018}
Popov, S.: The tangle - version 1.4.3 (2018),
  \url{http://web.archive.org/web/20190121105843/http://www.descryptions.com/Iota.pdf}

\bibitem{puddu_chain:_2017}
Puddu, I., Dmitrienko, A., Capkun, S.: uchain: How to forget without hard forks
  (2017), \url{https://eprint.iacr.org/2017/106.pdf}

\bibitem{puttgen_versicherung_2017}
P\"{u}ttgen, F., Kaulartz, M.: Versicherung 4.0. {ERA} Forum  \textbf{18}(2),
  249--262 (2017). \doi{10.1007/s12027-017-0479-y},
  \url{https://doi.org/10.1007/s12027-017-0479-y}

\bibitem{redman_segregated_2017}
Redman, J.: Segregated witness has officially activated on the bitcoin network
  (2017),
  \url{https://news.bitcoin.com/segregated-witness-has-officially-activated-on-the-bitcoin-network/}

\bibitem{altshuler_analysis_2013}
Reid, F., Harrigan, M.: An analysis of anonymity in the bitcoin system. In:
  Altshuler, Y., Elovici, Y., Cremers, A.B., Aharony, N., Pentland, A. (eds.)
  Security and Privacy in Social Networks, pp. 197--223. Springer New York
  (2013). \doi{10.1007/978-1-4614-4139-7\_10},
  \url{http://link.springer.com/10.1007/978-1-4614-4139-7\_10}

\bibitem{ren_bandwidth_2017}
Ren, L., Devadas, S.: Bandwidth hard functions for {ASIC} resistance. In:
  Theory of Cryptography. pp. 466--492. Lecture Notes in Computer Science,
  Springer International Publishing (2017)

\bibitem{rimba_comparing_2017}
Rimba, P., Tran, A.B., Weber, I., Staples, M., Ponomarev, A., Xu, X.: Comparing
  blockchain and cloud services for business process execution. In: 2017 IEEE
  International Conference on Software Architecture (ICSA). pp. 257--260.
  {IEEE} (2017). \doi{10.1109/ICSA.2017.44},
  \url{http://ieeexplore.ieee.org/document/7930226/}

\bibitem{noauthor_what_nodate}
Rosic, A.: What is ethereum classic? ethereum vs ethereum classic (2018),
  \url{https://blockgeeks.com/guides/what-is-ethereum-classic/}

\bibitem{ruffing_coinshuffle:_2014}
Ruffing, T., Moreno-Sanchez, P., Kate, A.: {CoinShuffle}: Practical
  decentralized coin mixing for bitcoin. In: Computer Security - {ESORICS}
  2014. pp. 345--364. Lecture Notes in Computer Science, Springer International
  Publishing (2014)

\bibitem{van_saberhagen_cryptonote_2013}
van Saberhagen, N.: {CryptoNote} v 2.0 (2013),
  \url{https://cryptonote.org/whitepaper.pdf}

\bibitem{sankar_survey_2017}
Sankar, L.S., Sindhu, M., Sethumadhavan, M.: Survey of consensus protocols on
  blockchain applications. In: 2017 4th International Conference on Advanced
  Computing and Communication Systems ({ICACCS}). pp.~1--5 (2017).
  \doi{10.1109/ICACCS.2017.8014672}

\bibitem{schiener_primer_2017}
Schiener, D.: A primer on {IOTA} (with presentation) (2017),
  \url{https://blog.iota.org/a-primer-on-iota-with-presentation-e0a6eb2cc621}

\bibitem{noauthor_raiden_2018}
Sheikh, J.: Raiden network 0.14.0 documentation (2018),
  \url{https://raiden-network.readthedocs.io/en/stable/}

\bibitem{silverstein_blockchain_nodate}
Silverstein, S., Cadigan, T.N.: A blockchain without cryptocurrency is just a
  database innovation - and that's great (2018),
  \url{https://www.businessinsider.com/blockchain-with-no-cryptocurrency-a-database-innovation-2018-2}

\bibitem{skudnov_mini-blockchain_2017}
Skudnov, R.: The mini-blockchain scheme (2017),
  \url{http://cryptonite.info/files/mbc-scheme-rev3.pdf}

\bibitem{sompolinsky_secure_2015}
Sompolinsky, Y., Zohar, A.: Secure high-rate transaction processing in bitcoin.
  In: Financial Cryptography and Data Security. pp. 507--527. Lecture Notes in
  Computer Science, Springer, Berlin, Heidelberg (2015).
  \doi{10.1007/978-3-662-47854-7\_32},
  \url{https://link.springer.com/chapter/10.1007/978-3-662-47854-7\_32}

\bibitem{swan_blockchain_2015}
Swan, M.: Blockchain thinking : The brain as a decentralized autonomous
  corporation. {IEEE} Technology and Society Magazine  \textbf{34}(4),  41--52
  (2015). \doi{10.1109/MTS.2015.2494358}

\bibitem{swan_blockchain:_2015}
Swan, M.: Blockchain: blueprint for a new economy. O'Reilly, 1th edition edn.
  (2015), {OCLC}: ocn898924255

\bibitem{szabo_bit_2008}
Szabo, N.: Bit gold (2008-12),
  \url{http://unenumerated.blogspot.com/2005/12/bit-gold.html}

\bibitem{tasatanattakool_blockchain:_2018}
Tasatanattakool, P., Techapanupreeda, C.: Blockchain: Challenges and
  applications. In: 2018 International Conference on Information Networking
  ({ICOIN}). pp. 473--475 (2018). \doi{10.1109/ICOIN.2018.8343163}

\bibitem{tschorsch_bitcoin_2016}
Tschorsch, F., Scheuermann, B.: Bitcoin and beyond: A technical survey on
  decentralized digital currencies. {IEEE} Communications Surveys \& Tutorials
  \textbf{18}(3),  2084--2123 (2016). \doi{10.1109/COMST.2016.2535718},
  \url{http://ieeexplore.ieee.org/document/7423672/}

\bibitem{underwood_blockchain_2016}
Underwood, S.: Blockchain beyond bitcoin. Communications of the {ACM}
  \textbf{59}(11),  15--17 (2016). \doi{10.1145/2994581},
  \url{http://doi.acm.org/10.1145/2994581}

\bibitem{vigna_bitcoin_2017}
Vigna, P.: Bitcoin dodges split that threatened its surging price. Wall Street
  Journal  (2017),
  \url{https://www.wsj.com/articles/bitcoin-dodges-split-that-threatened-its-surging-price-1510172701}

\bibitem{visa_inc._visa_nodate}
{Visa Inc.}: Visa inc. at a glance (2015),
  \url{https://usa.visa.com/dam/VCOM/download/corporate/media/visa-fact-sheet-Jun2015.pdf}

\bibitem{vorick_sia:_2014}
Vorick, D., Champine, L.: Sia: Simple decentralized storage (2014),
  \url{https://sia.tech/sia.pdf}

\bibitem{de_vries_bitcoins_2018}
de~Vries, A.: Bitcoin's growing energy problem. Joule  \textbf{2}(5),  801--805
  (2018). \doi{10.1016/j.joule.2018.04.016},
  \url{https://www.sciencedirect.com/science/article/pii/S2542435118301776}

\bibitem{vukolic_quest_2015}
Vukoli\'{c}, M.: The quest for scalable blockchain fabric: Proof-of-work vs.
  {BFT} replication. In: Open Problems in Network Security. pp. 112--125.
  Lecture Notes in Computer Science, Springer, Cham (2015).
  \doi{10.1007/978-3-319-39028-4\_9},
  \url{https://link.springer.com/chapter/10.1007/978-3-319-39028-4\_9}

\bibitem{walch_path_2017}
Walch, A.: The path of the blockchain lexicon (and the law). Review of Banking
  \& Financial Law  \textbf{36},  713--765 (2017),
  \url{https://papers.ssrn.com/abstract=2940335}

\bibitem{van_wegberg_bitcoin_2018}
van Wegberg, R., Oerlemans, J.J., van Deventer, O.: Bitcoin money laundering:
  mixed results? an explorative study on money laundering of cybercrime
  proceeds using bitcoin. Journal of Financial Crime pp. 00--00 (2018).
  \doi{10.1108/JFC-11-2016-0067},
  \url{http://www.emeraldinsight.com/doi/10.1108/JFC-11-2016-0067}

\bibitem{wessling_how_2018}
Wessling, F., Ehmke, C., Hesenius, M., Gruhn, V.: How much blockchain do you
  need?: Towards a concept for building hybrid {DApp} architectures. In:
  Proceedings of the 1st International Workshop on Emerging Trends in Software
  Engineering for Blockchain. pp. 44--47. {WETSEB} '18, {ACM} (2018).
  \doi{10.1145/3194113.3194121},
  \url{http://doi.acm.org/10.1145/3194113.3194121}

\bibitem{wessling_tactics_2019}
Wessling, F., Ehmke, C., Meyer, O., Gruhn, V.: Towards blockchain tactics:
  Building hybrid decentralized software architectures. In: 2019 {IEEE}
  International Conference on Software Architecture Companion ({ICSA}-C) (2019)

\bibitem{wilkinson_storj_2016}
Wilkinson, S., Boshevski, T., Brandoff, J., Prestwich, J., Hall, G., Gerbes,
  P., Hutchins, P., Pollard, C.: Storj a peer-to-peer cloud storage network
  (2016), \url{https://storj.io/storj.pdf}

\bibitem{willett_second_2012}
Willett, J.R.: The second bitcoin whitepaper (2012),
  \url{https://sites.google.com/site/2ndbtcwpaper/2ndBitcoinWhitepaper.pdf}

\bibitem{wilmoth_bitmains_2018}
Wilmoth, J.: Bitmain's mining pools now control nearly 51\% of the bitcoin
  hashrate (2018),
  \url{https://www.ccn.com/bitmains-mining-pools-now-control-nearly-51-percent-of-the-bitcoin-hashrate/}

\bibitem{wood_ethereum:_2014}
Wood, G.: Ethereum: A secure decentralised generalised transaction ledger
  (2014), \url{http://gavwood.com/Paper.pdf}

\bibitem{wust_you_2017}
W\"ust, K., Gervais, A.: Do you need a blockchain? (2017),
  \url{http://eprint.iacr.org/2017/375}

\bibitem{xu_taxonomy_2017}
Xu, X., Weber, I., Staples, M., Zhu, L., Bosch, J., Bass, L., Pautasso, C.,
  Rimba, P.: A taxonomy of blockchain-based systems for architecture design.
  In: 2017 {IEEE} International Conference on Software Architecture ({ICSA}).
  pp. 243--252 (2017). \doi{10.1109/ICSA.2017.33}

\bibitem{yaga_blockchain_2018}
Yaga, D., Mell, P., Roby, N., Scarfone, K.: Blockchain technology overview.
  Tech. Rep. {NIST} {IR} 8202, National Institute of Standards and Technology
  (2018). \doi{10.6028/NIST.IR.8202},
  \url{https://nvlpubs.nist.gov/nistpubs/ir/2018/NIST.IR.8202.pdf}

\bibitem{yelowitz_characteristics_2015}
Yelowitz, A., Wilson, M.: Characteristics of bitcoin users: an analysis of
  google search data. Applied Economics Letters  \textbf{22}(13),  1030--1036
  (2015). \doi{10.1080/13504851.2014.995359},
  \url{https://doi.org/10.1080/13504851.2014.995359}

\bibitem{zamanov_asic-resistant_2018}
Zamanov, A.R., Erokhin, V.A., Fedotov, P.S.: {ASIC}-resistant hash functions.
  In: 2018 {IEEE} Conference of Russian Young Researchers in Electrical and
  Electronic Engineering ({EIConRus}). pp. 394--396 (2018).
  \doi{10.1109/EIConRus.2018.8317115}

\bibitem{zhang_towards_2018}
Zhang, K., Jacobsen, H.A.: Towards dependable, scalable, and pervasive
  distributed ledgers with blockchains. In: 2018 {IEEE} 38th International
  Conference on Distributed Computing Systems ({ICDCS}). pp. 1337--1346 (2018).
  \doi{10.1109/ICDCS.2018.00134}

\bibitem{zheng_overview_2017}
Zheng, Z., Xie, S., Dai, H., Chen, X., Wang, H.: An overview of blockchain
  technology: Architecture, consensus, and future trends. In: 2017 {IEEE}
  International Congress on Big Data ({BigData} Congress). pp. 557--564 (2017).
  \doi{10.1109/BigDataCongress.2017.85}

\end{thebibliography}

\end{document}